\documentclass[12pt]{article}
\usepackage{hyperref}
\usepackage{amssymb}
\usepackage{xcolor}
\usepackage[mathscr]{eucal}
\usepackage{url}
\usepackage{hyperref}
\usepackage{a4}
\usepackage{cite}
\usepackage{graphicx}
\usepackage{psfrag}
\usepackage{subfigure}
\usepackage{latexsym}
\usepackage{amsmath}
\usepackage[utf8]{inputenc} 
\usepackage{cancel}

\newcommand{\xitau}{\red{\omega}}
\newcommand{\myxitau}{\xitau}
\newcommand{\myxi}{\red{k}}

\newcommand{\mysigma}{\blue{\sigma}}
\newcommand{\bundle}{\blue{F}}

\newcommand{\mcO}{\blue{\mycal O}}

\newcommand{\wancheck}[1]{\mnote{wan: checked on #1}}

\newcounter{mnotecount}[section]
\renewcommand{\themnotecount}{\thesection.\arabic{mnotecount}}
\newcommand{\mnote}[1]
{\protect{\stepcounter{mnotecount}}$^{\mbox{\footnotesize $
\bullet$\themnotecount}}$ \marginpar{
\raggedright\tiny\em $\!\!\!\!\!\!\,\bullet$\themnotecount: #1} }

\DeclareFontFamily{OT1}{rsfs}{}
\DeclareFontShape{OT1}{rsfs}{m}{n}{ <-7> rsfs5 <7-10> rsfs7 <10-> rsfs10}{}
\DeclareMathAlphabet{\mycal}{OT1}{rsfs}{m}{n}

\newcommand{\zh}{\mathring{h}}

\newcommand{\mcV}{\blue{\mycal{V}}}
\newcommand{\X}{\blue{\mycal{X}}}

\newtheorem{theorem}{\sc  Theorem\rm}[section]

\newtheorem{Definition}[theorem]{\sc  Definition\rm}

\newtheorem{Proposition}[theorem]{\sc Proposition\rm}
\newtheorem{remark}[theorem]{\sc Remark\rm}

\newtheorem{Remarks}[theorem]{\sc Remarks\rm}

\newcommand{\mcU}{\blue{\mycal{U}}}
\newcommand{\zD}{\blue{\mathring{D}}}
\newcommand{\nozD}{D}

\newcommand{\loc}{\blue{\mathrm{loc}}}

\newcommand{\mcD}{\blue{\mycal D}}

\newcommand{\mcL}{\blue{\mycal{L}}}

\newcommand{\mychi}{\blue{\chi}}

\newcommand{\cC}{\blue{ \mathcal{C}}}
\newcommand{\mcC}{\cC}

\newcommand{\red}[1]{{\color{red}#1}}

\newcommand{\ptcheck}[1]{\ptcr{checked on #1}}

\newcommand{\blue}[1]{{\color{blue}#1}}

\newcommand{\Gammapp}{\blue{\Gamma_p^+}}
\newcommand{\mcTpp}{\blue{\mycal{T}_p^+}}
\newcommand{\mcTp}{\blue{\mycal{T}_p}}

\newcommand{\mcNp}{\blue{\mycal{N}_p}}

\newcommand{\mcM}{\mycal{M}}

\newcommand{\hyp}{\mycal{S}}

\newcommand{\fourg}{{\mathfrak g }}

\newcommand{\R}{\mathbb R}

\newcommand{\N}{\mathbb N}

\newcommand{\ptcr}[1]{\mnote{\color{red}ptc: #1}}

{\catcode `\@=11 \global\let\AddToReset=\@addtoreset}
\AddToReset{equation}{section}

{\catcode `\@=11 \global\let\AddToReset=\@addtoreset}
\AddToReset{figure}{section}

{\catcode `\@=11 \global\let\AddToReset=\@addtoreset}
\AddToReset{table}{section}

\renewcommand{\ptcheck}[1]{}
\renewcommand{\blue}[1]{#1}

\renewcommand{\red}[1]{#1}

\renewcommand{\wancheck}[1]{}

\begin{document}
\newcommand{\g}{$\bf \bar g g^{\alpha\beta}$}
\title{Who's afraid of a negative lapse?}

 \author{Robert Beig\footnote{Faculty of Physics, University of  Vienna, Austria},
Piotr T. Chru\'{s}ciel\footnote{Beijing Institute for Mathematical Sciences and Applications, and
Center for Theoretical Physics, Polish Academy of Sciences, Warsaw}, and 
Wan Cong$^*$
}

\maketitle

\begin{abstract} 
We rederive the Arnowitt-Deser-Misner equations in a framework in which the zeros of the lapse are innocuous, whether with or without changes of sign. 
We further develop and analyse a covariantized version of  the Anderson-York   equations, which  provide a well posed system of tensorial  evolution equations with freely prescribable shift vector and densitised lapse.  
The causality properties of the resulting equations  are explored.
We show how to relate solutions of the Anderson-York equations to the maximal globally hyperbolic development of the initial data.
\end{abstract}

\tableofcontents


\section{Introduction}

One way to cast general relativity  into the form of a dynamical system proceeds through the Arnowitt-Deser-Misner  (ADM)  parameterisation of the metric, in which we have
\begin{equation}\label{ADM0}
 g= 
 - N^2 d\tau^2 + h_{ij} (dy^i + X^i d \tau) (dy^j + X^j d \tau)
  \,,
\end{equation}
and where one assumes that $N$ has no zeros to guarantee a Lorentzian signature. Given a lapse function $N$ and a shift vector $X^i$, in vacuum  one has the following first-order evolution system for a pair of fields $(h_{ij},K_{ij})$: 
\begin{equation}\label{Dvh'a}
(\partial_\tau - \mathcal{L}_X) h_{ij} = 2 N K_{ij}
\,,
\end{equation}
\begin{equation}\label{DvKa}
(\partial_\tau - \mathcal{L}_X) K_{ij} = - N (\mathcal{R}_{ij} - \lambda h_{ij}) + 2 N 
 K_{i\ell} K^\ell{}_j  - NK K_{ij} + D_i D_j N
 \,,
\end{equation}
with $K= K^i{}_i$, where $\mathcal{R}_{ij}$ is the Ricci tensor of the metric $h_{ij}$ and $\lambda$ is proportional to the cosmological constant $\Lambda $:
in spacetime dimension $n+1$,
$$
 \lambda = \frac{2 \Lambda  }{n-1}
 \,.
$$
These are complemented by the constraints:
\begin{equation}\label{C,Cia}
\mathcal{C} =: \frac{1}{2}(\mathcal{R} - 2 \Lambda  - K_{ij} K^{ij} + K^2) = 0\,,\hspace{1cm}\mathcal{C}_i := D_j(K_i{}^j - \delta_i{}^j K) = 0
 \,.
\end{equation}
Equations (\ref{Dvh'a})-(\ref{DvKa}), for given lapse and shift, form a determined system of evolution equations for $(h_{ij}, K_{ij})$ with initial data subject to the constraints (\ref{C,Cia}). The evolution system (again for given $(N, X^i)$) is first order in `time' and second order in space derivatives due to the presence of the Ricci tensor.  

It has been discovered by Anderson and York~\cite{anderson1999fixing} that these equations are part of a  well-posed symmetric hyperbolic system of equations when $N$ is replaced by a freely-prescribable field 
\begin{equation}\label{4II25.2-1}
  Q := (\det h)^{-1/2} N
  \,.
\end{equation}
The resulting system of equations is well posed even if $Q$ changes sign.  
 
Note that $Q=\text{const}$ corresponds to a time-coordinate which satisfies the massless wave equation when the shift vanishes.

Alternative symmetric hyperbolic rewritings of the ADM equations for given $(Q',X^i)$ with $Q'$ some 'densitisation' of $N$ exist, see~\cite{kidder2001extending} for an overview. A well posed system with lapse $N=1$ and zero shift has been employed
in~\cite{FournodavlosLuk,fournodavlos2021initial}.  
There exists a vast literature on hyperbolic formulations using a dynamical lapse and shift (see \cite{Hilditch:2024nhf} and references therein),  mostly based on the BSSN-formulation of the ADM equations. 
 
The first main 
 aim of  this work is to  
present a formalism which seamlessly accomodates zeros of the lapse function. We use this to  provide a derivation, first of equations (\ref{Dvh'a})-(\ref{DvKa}), and then of the Anderson-York  evolution equations, in a way    which makes it clear that zeros of $Q$ are innocuous in the problem at hand.%
\footnote{See, e.g., \cite{baumgarte2022shock} for examples of numerical evolution where $N$ changes sign.}
The idea is to view the evolution  as a {spacelike slicing} of spacetime, 
as in~\cite{CJK} (see also~\cite{PSchmidt}), 
where the dynamics of the gravitational field
 is associated with a one-parameter family of spacelike slices of the spacetime.

A key element of any treatment of the general Cauchy problem is the proof of preservation of constraints. The original paper~\cite{anderson1999fixing} glosses over this%
\footnote{The authors of~\cite{anderson1999fixing} note that  ``A mathematically well-posed form of
the twice-contracted Bianchi identities [27,28] shows
that these initial-value constraints remain satisfied if the
equations of motion are equivalent to $R_{ij}= 8\pi T_{ij} -
1/2 g_{ij} T^\mu{}_\mu$.''  (refs. as in op.\ cit.). But this latter equivalence does not hold before one has shown that all constraints are propagated, regardless of whether or not $N$ has  zeros, see \eqref{4II25.3b} below.}
, without addressing the fact that the usual general relativistic constraints get coupled with the constraints introduced by the rewriting of the equations as a first order system, while using the vector constraint equation in the process. 
This leads to a system for the propagation of constraints different from the one hinted-to in~\cite{anderson1999fixing}. 
In any case, we settle this question in the affirmative
(compare~\cite{kidder2001extending}; 
see Section~\ref{s18VIII25.1} for terminology):

\begin{theorem}
  \label{T1VI25.1}
  The system of constraints associated with the {Anderson-York} equations is microlocally symmetrisable-hyperbolic. 
\end{theorem} 

It follows~\cite{Taylor91}  that
the {Anderson-York} equations  
(referred to as \emph{Einstein-Christoffel equations} in \cite{kidder2001extending,anderson1999fixing}) 
can be used to provide  solutions of the Einstein equations with initial data in Sobolev spaces, which was neither clear from~\cite{anderson1999fixing}, nor from the remarks on the constraint propagation in~\cite{kidder2001extending}. 

While the tensor field \eqref{ADM0} degenerates at zeros of $Q$, hence of $N$, the evolution equations remain regular, with smooth local solutions for smooth initial data $(h_{ij},K_{ij})|_{\tau=0}$ with a Riemannian $h_{ij}|_{\tau=0}$. The solutions are unique in domains of dependence as defined in Section~\ref{s21I25.1}. This allows us to prove the second main result of this paper:

\begin{theorem}
  \label{t!VI25.1}
 Globally hyperbolic,  in the sense of Section~\ref{s21I25.1}, solutions of the Anderson-York equations    can be mapped into the usual maximal globally hyperbolic development of the initial data.
\end{theorem}

A  more precise statement   can be found in Theorem~\ref{T2VI25.1} below. 

The result holds again regardless of zeros of $Q$, so that no causal inconsistencies can arise, even when the evolution via the Anderson-York equations passes several times through the same event in the maximal globally hyperbolic development of the initial data.

Theorem~\ref{t!VI25.1} is rather clear when $Q$ has no zeros, a formal proof in this last case is presented in Section~\ref{s8IV25.1}, see Proposition~\ref{P2VI26.1}. 
The general case  is established  in Section~\ref{s13IV25.1}, as part of the solution of the \emph{slicing problem} analysed there.

Indeed, given a spacetime $(\mcM,g)$ and a spacelike hypersurface $\hyp$, whether vacuum or not, we formulate the  \emph{ $(Q,X)$ slicing problem} as follows: given a spacetime $(\mcM,g)$, a densitised lapse $Q$, and a shift vector $X$, can 
one find a slicing in $\mcM$ which realises $Q$ and $X$?
The third main result of this work is the following answer to this question:

\begin{theorem}
  \label{T29V25.1}
  Given smooth fields $(Q,X^i)$,
the slicing problem  can be solved locally.
\end{theorem}

\section{Slices}
 \label{s29III25.1}

We define a  
\emph{spacelike slice} 
as the image of  an embedding  
$$
 \phi: \hyp \mapsto M
$$
of a 
$n$-dimensional manifold $\hyp$, with $n \geq 2$, 
 into a Lorentzian manifold $(M, g_{\mu\nu})$ of dimension $n+1$. In local coordinates $y^i$ on $\hyp$ and $x^\mu$ on $M$ this means that the metric on $\hyp$, given by
\begin{equation}
h_{ij}(y) = \phi^\mu{}_{{, i}}(y) \phi^\nu{}_{{, j}}(y) g_{\mu\nu}(\phi(y))
 \,,
\end{equation}
is positive definite. It follows that there exists a one-form $n_\mu (y)$ defined uniquely up-to-sign  by the equations
\begin{equation}
\phi^\mu{}_{{, i}} (y) n_\mu (y) = 0\,,\hspace{1cm}g^{\mu\nu}(\phi(y)) n_\mu(y) n_\nu (y) = - 1
\,.
\end{equation}
We choose $n_\mu$ to be such that $n^\mu = g^{\mu\nu}n_\nu$ is future-pointing relative to some time orientation of $M$ and call this the unit normal. For example, in spacetime dimension $n+1=3+1$, the field $n_\mu$ can be explicitly written in terms of $\phi$ as 
\begin{equation}
 \label{4V25.1}
n_\mu = \frac{1}{3!} \phi^\nu{}_{{, i}}\phi^\rho{}_{{, j}}\phi^\sigma{}_{{, k}}\epsilon^{ijk}\epsilon_{\mu\nu\rho\sigma}\,,
\end{equation}
where $\epsilon_{ijk}$ is the volume element of $h_{ij}$.

For  later use  we note the  following `completeness relation':
\begin{equation}\label{compl}
\phi^\mu{}_{{, i}} \phi^\nu{}_{{, j}} h^{ij} = g^{\mu\nu} + n^\mu n^\nu = : h^{\mu\nu}\,.
\end{equation}
We will also use the field $\psi^i{}_\mu$  uniquely defined by the conditions
  $\psi^i{}_\mu n^\mu = 0$ and
\begin{equation}
\phi^\mu{}_{{, i}} \psi^i{}_\nu = h^\mu{}_\nu\,,\hspace{1cm} \phi^\mu{}_{{, j}}\psi^i{}_\mu = \delta^i{}_j
 \,.
\end{equation}
The field $\psi^i{}_\mu$ can be explicitly written as
\begin{equation}\label{psi}
\psi^i{}_\mu = \phi^\nu{}_{{, j}} g_{\mu\nu} h^{ij}\,.
\end{equation}
The objects appearing in the above equations are all functions of $y \in \hyp$. 
 Some, such as $h_{ij}$, are tensor fields on $\hyp$. Others, such as $g^{\mu\nu}$, are tensor fields on $M$, evaluated at points $x = \phi(y)$. 
 These are special cases of the concept of \emph{two-point tensors} over a map $\phi$. These are objects, say $t_{i...\mu..}{}^{j..\nu..}(y)$, which are simultaneously tensor fields on $\hyp$ and  tensors on $M$ at points $x = \phi(y)$ (see
\cite{marsden1994mathematical}, which refers to  
\cite[Appendix. Tensor Fields. by J.L. Ericksen]{ericksen1960appendices}, which in turn attributes the concept back to Clebsch in 1872 \cite{Clebsch}. 
In modern language: Let $\mathcal{S}$, respectively $\mathscr{T}$, be tensor bundles over $\hyp$, respectively $M$. 
Then a two-point tensor is a cross section of the product bundle $\mathcal{S} \otimes \phi^\star \mathscr{T}$ over $\hyp$. The concept also appears in treatments of harmonic maps, see e.g.
\cite{eells1964harmonic}%
\footnote{But apparently the only place where this concept is used in a systematic fashion is the field of continuum mechanics, see e.g.\ the notion of first Piola stress tensor.}.
A prime example in our context, in addition to the above,  is the field $n_\mu(y)$, which is a scalar on $\hyp$ and a covector at $\phi(y)$. Another one is  $\phi^\mu{}_{{, i}}(y)$,
which is a covector field with respect to $\hyp$ and a vector field with respect to $M$. Thus, under changes of coordinates this last field behaves as
\begin{equation}
\bar{V}^{\mu'}{}_{i'} (\bar{y}) = \frac{\partial \bar{x}^{\mu'}}{\partial x^\nu}(\phi(y(\bar{y})))\frac{\partial y^{i}}{\partial \bar{y}^{i'}}(\bar{y})
V^\nu{}_i(y(\bar{y}))\,.
\end{equation}
We will denote by  $\nabla_\mu$ 
the Levi-Civita connection of $(M, g_{\mu\nu})$ and by  $D_i$ the Levi-Civita connection 
of $(\hyp,h_{ij})$. 
We can extend $D_i$ to two-point tensors in the obvious way. For example,  the derivative $D_i$ of 
a two-point tensor of type $V^i{}_\mu$ is defined as 
\begin{equation}
D_j V^i{}_\mu = \partial_j V^i{}_\mu - \Gamma^\nu_{\mu\rho} V^i{}_\nu \phi^\rho{}_{{, j}} + \gamma^i_{jk} V^k{}_\mu\,,
\end{equation}
where it is understood that the Christoffel symbols $\Gamma^\nu_{\mu\rho}$ of $g_{\mu\nu}$ are evaluated at points $x = \phi(y)$, and where
 $\gamma^k_{ij}$ denotes the Christoffel symbols of $h_{ij}$. This operation, when extended in the 
  usual way to two-point tensors of all types, satisfies the familiar rules with respect to tensor products and contractions on both $\hyp$ and $M$. Furthermore, when acting on tensor fields on $M$ at $x = \phi(y)$, it coincides with $\phi^\mu{}_{{, i}} \nabla_\mu$ 
  and when acting on purely spatial tensors it coincides with the standard covariant derivative on $\hyp$.
 
It is not difficult to prove that 
\begin{equation}\label{curv}
[D_{i}, D_{j}]V^\mu{}_k = \phi^\rho{}_{{, i}}\phi^\sigma{}_{{, j}}R_{\rho\sigma}{}^\mu{}_\nu V^\nu{}_k + \mathcal{R}_{ijk}{}^l V^\mu{}_l\,,
\end{equation}
where $R_{\rho\sigma}{}^\mu{}_\nu$ is the curvature tensor of $g_{\mu\nu} $ and $\mathcal{R}_{ijk}{}^l$ is that of $h_{ij}$.

An important two-point tensor is the \emph{extrinsic curvature} tensor defined as
\begin{equation}\label{nons}
K^\mu{}_{ij} = K^\mu{}_{ji} = D_i \phi^\mu{}_{{, j}} = \phi^\mu{}_{{, i}j} + \Gamma^\mu_{\nu\rho} \phi^\nu{}_{{, i}}\phi^\rho{}_{{, j}} - \gamma^l_{ij} \phi^\mu{}_{{, l}}\,.
\end{equation}
%
A useful identity relating $\nabla$ and $D$ is
\begin{multline}\label{newid}
 \phi^\mu{}_{{, i}}\cdots\phi^\nu{}_{{, j}}\phi^\rho{}_{{, l}}
  \nabla_\rho V_{\mu\cdots\nu} =\phi^\mu{}_{{, i}}\cdots\phi^\nu{}_{{, j}} D_l V_{\mu\cdots\nu} = D_l (\phi^\mu{}_{{, i}}\cdots\phi^\nu{}_{{, j}}V_{\mu\cdots\nu}) - \\
- K^\mu{}_{il}\phi^\nu{}_{{, j}}V_{\mu\cdots\nu}\,\,\,...-\phi^\mu{}_{{, i}}\,...K^\nu{}_{jl}V_{\mu\cdots\nu}\,.
\end{multline}
So far we have not used the fact that $h_{ij}$ is the pull-back of $g_{\mu\nu}$ under $\phi$, 
whence $D_i$ annihilates the induced metric $h_{ij}$.  
Doing this we find that  
\begin{equation}\label{Dhij}
D_i h_{jk} = D_i (\phi^\mu{}_{{, j}}\phi^\nu{}_{{, k}}\, g_{\mu\nu}) = 2 K^\mu{}_{i(j}\phi^\nu{}_{{, k})}\, g_{\mu\nu} = 0\hspace{0.2cm}\Rightarrow\hspace{0.2cm}
K^\mu{}_{ij}\phi^\nu{}_{{, k}}\, g_{\mu\nu} = 0\,,
\end{equation}
where we used $K^\mu{}_{ij} = K^\mu{}_{(ij)}$.
Thus there is a tensor $K_{ij} = K_{(ij)}$ such that
\begin{equation}\label{Dphi}
K^\mu{}_{ij} = n^\mu K_{ij}\,.
\end{equation}
Contracting this equation with $n_\mu$ and using (\ref{nons}) we see that $K_{ij}$ can also be defined by
\begin{equation}\label{standard}
D_i n_\mu = \psi^j{}_\mu K_{ij}\,.
\end{equation}
We will find both definitions of the extrinsic curvature useful.

Applying the identity (\ref{curv}) to $\phi^\mu{}_{{, k}}$ and using (\ref{nons}), (\ref{Dphi}) and (\ref{standard}) there results in
\begin{equation}\label{R+R}
2\, \phi^\mu{}_{{, l}} K^l{}_{[i} K_{j]k} + 2\, n^\mu D_{[i}K_{j]k} = \phi^\rho{}_{{, i}}\phi^\sigma{}_{{, j}}R_{\rho\sigma}{}^\mu{}_\nu \phi^\nu{}_{{, k}} + \mathcal{R}_{ijk}{}^l \phi^\mu{}_{{, l}}\,,
\end{equation}
which entails both the Codazzi
\begin{equation}\label{Cod}
R_{ijl\mu} n^\mu := \phi^\mu{}_{{, i}} \phi^\nu{}_{{, j}}\phi^\rho{}_{{, l}} R_{\mu\nu\rho\sigma} n^\sigma = 2 D_{[i} K_{j]l}
\end{equation}
and the Gauss equation
\begin{equation}\label{Gauss}
R_{ijlm} := \phi^\mu{}_{{, i}} \phi^\nu{}_{{, j}}\phi^\rho{}_{{, l}} \phi^\sigma{}_{,m} R_{\mu\nu\rho\sigma} = \mathcal{R}_{ijlm} - 2 K_{m[i} K_{j]l}\,.
\end{equation}
These imply (recall that 
$K= h^{ij}K_{ij}$)
the usual vacuum constraint equations
\begin{eqnarray}\label{CCia}
 & 
\mathcal{C}_i :=  
 \phi^\mu{}_{,i} (G_{\mu\nu} + \Lambda  g_{\mu\nu}) n^\nu = D_j(K_i{}^j - \delta_i{}^j K)=0
\,,
&
\\
& 
\mathcal{C} := (G_{\mu\nu} +\Lambda g_{\mu\nu} ) n^\mu n^\nu = \frac{1}{2}(\mathcal{R} - 2 \Lambda  - K_{ij} K^{ij} + K^2) = 0
 \,.
 &
 \label{CCib}
\end{eqnarray}

\section{Slicings}
 \label{s31I25.1}

By a spacelike slicing we mean a smooth one-parameter family of differentiable spacelike embeddings, 
 i.e.\
  maps $(\tau \in (- \epsilon, \epsilon),y \in \hyp) \mapsto \phi^\mu(\tau,y) \in M$, where for each $\tau$ the image of $\hyp$ by  $\phi(\tau,\cdot)$ is an $n$-dimensional spacelike embedded  hypersurface  in an  $(n+1)$-dimensional spacetime $M$. 
  
We can decompose $\partial_\tau \phi = : \dot{\phi}^\mu$ as
\begin{equation}\label{lapse-shift}
\dot{\phi}^\mu (\tau,y) = N (\tau,y) n^\mu (\tau,y) + \phi^\mu{}_{{, i}} (\tau,y) X^i (\tau,y)
 \,.
\end{equation}
The fields $N$ and $X^i$ are called the \emph{lapse function} and the \emph{shift vector field} of the  slicing. 
A useful form of rewriting (\ref{lapse-shift}) is
\begin{equation}\label{useful}
(\partial_\tau - X^\ell\partial_\ell ) \phi^\mu = N n^\mu\,.
\end{equation}

When the lapse  has no zeros, $\phi^\mu (\tau,y)$ maps $(- \epsilon, \epsilon) \times \hyp$ diffeomorphically onto its image and
defines $\tau(x^\mu)$ as a time function associated with a foliation of $M$ by spacelike hypersurfaces. Then the normal $n_\mu$ can be viewed as a covector field on $M$, as is done in usual treatments which are based on a time function. For non-vanishing lapse, formulae such as $K=\nabla_\mu n^\mu$ involving the spacetime divergence are perfectly legitimate, but do  not make sense for general slicings, where e.g.\
$$
 K =  \psi^i{}_\mu{} D_i n^\mu
$$
can be used instead.

What always makes sense is the pull-back, which we denote by $(\cdot)^\star$, of covariant tensors on $M$ to
$(- \epsilon, \epsilon)\times \hyp$. Using
\begin{equation}
(dx^\mu)^\star = d \phi^\mu 
=\dot \phi^\mu d\tau + \phi^\mu{}_{{, i}}dy^i 
= N n^\mu d\tau + \phi^\mu{}_{{, i}}(dy^i + X^i d\tau)\,,
\end{equation}
we can write
\begin{equation}\label{ADM}
(g_{\mu\nu} dx^\mu dx^\nu)^\star = - N^2 d\tau^2 + h_{ij} (dy^i + X^i d \tau) (dy^j + X^j d \tau)\,.
\end{equation}
Using this notation it holds that
\begin{equation}
\mathrm{det}\, g^\star = - N^2\, \mathrm{det} \,h\,.
\end{equation}
Thus, when $h$ is Riemannian the pulled-back metric is Lorentzian, in particular non-degenerate, if and only if  the lapse $N$ is everywhere non-zero. This provides an easy proof that $(\tau, y^i)$ are good coordinates on Lorentzian $(M,g)$'s if and only if  $N$ is non-zero. Equation (\ref{ADM}) also shows that the vector field $\partial_\tau - X^\ell\partial_\ell $  is orthogonal to the vector fields $\partial_i$ on $\{\tau\} \times \hyp$ with respect to the pulled-back metric, and the former vector field annihilates this metric at a point $p$ if and only if  $N$ vanishes at $p$.

We wish to write
 the Einstein equations as evolution equations on $\{\tau\} \times \hyp$ with $\tau$ playing the role of ``time''. For that purpose we need a concept of time derivative on $\{\tau\} \times \hyp$ acting on two-point tensors, which is the ``vertical'' counterpart, which we denote by  $D_v$, of the \emph{horizontal derivative} $D_i$. For a two-point tensor of the type $t^\mu$ this derivative is defined as
\begin{equation}\label{vert}
D_v t^\mu = (\partial_\tau - X^\ell\partial_\ell )\, t^\mu + N \Gamma^\mu_{\!\nu\rho}n^\nu t^\rho \,.
\end{equation}
To check the tensorial nature of (\ref{vert}), 
evaluating $t^\mu$ at $x = \phi(y)$
and using  (\ref{useful}), we simply have $D_v t^\mu = N n^\nu \nabla_\nu t^\mu$. 

Next, in the presence of an additional index $i$, we define
\begin{equation}
D_v t^\mu{}_i = \partial_\tau t^\mu{}_i - X^\ell\partial_\ell  t^\mu{}_i + N \Gamma^\mu_{\nu\rho}n^\nu t^\rho{}_i - t^\mu{}_l \partial_i X^l
\end{equation}
One checks that the first three terms together, and the last term separately, behave correctly under $x^\mu$-transformations, while the first and third term separately, and the second and fourth term together, behave correctly under
transformations of the  $y^i$ coordinates on each slice.
Finally the definition (\ref{vert}), extended in the standard way to all types of two-point tensors, satisfies the usual rules for tensor products and contractions. Furthermore it coincides with $N n^\mu \nabla_\mu$ when acting on tensors on $M$,
and with the operator $\partial_\tau-\mathcal{L}_X$ when acting on quantities with Latin indices only. 

In what follows we will need some identities involving $D_v$. These are
\begin{equation}\label{Dvpsi}
D_v \phi^\mu{}_{{, i}} = D_i (N n^\mu) = (D_i N) n^\mu + N \phi^\mu{}_{{, l}}K_i{}^l
\end{equation}
and
\begin{equation}\label{Dvn}
D_v n_\mu = \psi^i{}_\mu D_i N\,.
\end{equation}
The identity (\ref{Dvpsi}) follows easily from
\begin{equation}
D_v \phi^\mu{}_{{, i}} = (\partial_\tau-X^\ell\partial_\ell ) \partial_i \phi^\mu + \Gamma^\mu_{\nu\rho} N n^\nu \phi^\rho{}_{{, i}} - \phi^\mu{}_{{, l}}\partial_i X^l\,.
\end{equation}
Commuting derivatives in the first expression on the right, the $\partial X$--terms cancel and using (\ref{useful}) we obtain (\ref{Dvpsi}).                         Next, the identity (\ref{Dvn}) follows from the definition of $n_\mu$, from Equation (\ref{Dvpsi}) and from the fact that $D_v g_{\mu\nu} = 0$. 

We finish this section with two commutator identities. First, 
for any covector field $V_\mu$ it holds that
\begin{equation}\label{id1}
[D_v, D_i] V_\mu = R_{\rho\sigma\mu}{}^\nu N n^\rho \phi^\sigma{}_{{, i}} V_\nu
 \,.
\end{equation}
In the absence of a more conceptual approach, this can be checked explicitly by a somewhat lengthy but completely straightforward calculation.  
This will be used in the calculations below with $V_\mu=n_\mu$. 

Finally, for any   covector field $W_j$,   
\begin{equation}\label{id2}
[D_v, D_i] W_j = - [D_i (N K_{jl}) + D_j (NK_{il}) - D_l(NK_{ij})] W^l
 \,.
\end{equation}
This is again checked by a lengthy calculation, and will not be used in what follows.
 
 \section{To shift or not to shift} 
 \label{s19VI25.1}
 
It is sometimes convenient to have the shift vector field $X^i$  around, therefore we have been carrying it along so far, and we will keep doing this in what follows. However, $X^i$ is essentially irrelevant for all local considerations 
in the ADM approach, and is irrelevant for all global ones if the orbits of $X^i$ are defined for the ranges of the parameter $\tau$ of interest. This is made precise by the considerations that  follow.

 Consider a slicing with a shift vector field $X^i$, and let $\phi_\tau[X]$ denote the (possibly local) flow generated by $X$. Let
\begin{align}\label{ADM19VII25.1}
  \hat N(\tau,y ) 
  \ & 
  : = (\phi_\tau[X]^*N)(\tau,y) \equiv N(\tau, \phi_\tau[X](y))
  \,, 
\\   
  \hat h_{ij}(\tau,y )
    \ &
    := \big(\phi_\tau[X]^* h(\tau, \cdot)\big)_{ij}(\tau,y)
  \,, 
\\
  \hat K_{ij}(\tau,y ) 
    \ &
    := \big(\phi_\tau[X]^* K(\tau, \cdot)\big)_{ij}(\tau,y)
   \,.
   \label{ADM19VII25.3}
\end{align}
It then follows from the definition of Lie derivative $\mathcal{L}_X$ in the first equality below, and from \eqref{Dvh'a} in the second, that 
\begin{align}\label{ADM19VII25.4} 
 \frac{\partial \hat h_{ij}(\tau,y )}{\partial \tau}
    \ &
    = \Big(\phi_\tau[X]^* 
    \big[
    \frac{\partial  h}{\partial \tau}+ \mathcal{L}_X h 
    \big]
    \Big) _{ij}(\tau,y)  
  \nonumber
\\
    \ &
    = \Big(\phi_\tau[X]^* 
    \big[
    2 N K 
    \big]
    \Big)_{ij}(\tau,y) 
  \,,
  \nonumber
\\
    \ &
    = 
  2 \hat N \hat K_{ij}(\tau,y )  
   \,.
\end{align}
A similar calculation applies to \eqref{DvKa}. We conclude that the fields defined in \eqref{ADM19VII25.1}-\eqref{ADM19VII25.3} satisfy the hatted equivalent of \eqref{Dvh'a}-\eqref{DvKa} with vanishing shift-vector.

In other words, one can get rid of $X^i$, or introduce it, or change it,
 by applying a flow to the slicing.  

As an application of this argument  we see that the proof~\cite{FournodavlosLuk} of well-posedness of the Einstein equations using the $(N=1,X^i=0)$-slicing generalises  to a $(N=1,X^i)$-slicing, with an arbitrary shift vector $X^i$, 
provided that the orbits of $X^i$ are defined for the time needed for the remaining considerations of~\cite{FournodavlosLuk}.

\section{The evolution equations}
 \label{s9III25.1}
 
\subsection{The extrinsic curvature tensor}
 \label{ss9III25.1}
  
First compute
\begin{equation}\label{Dvh}
(\partial_\tau - \mathcal{L}_X) h_{ij} = D_v(\phi^\mu{}_{{, i}}\phi^\nu{}_{{, j}} g_{\mu\nu}) = 2 (n^\mu D_i N + N \phi^\mu{}_{{, l}}K_i{}^l) \phi^\nu{}_{{, j}} g_{\mu\nu} = 2 N K_{ij}
 \,,
\end{equation}
where we have used (\ref{Dvpsi}). 

Equation~\eqref{Dvh}, together with (\ref{psi}), (\ref{Dvpsi}), and (\ref{Dvh}) in the form
\begin{equation}
(\partial_\tau - \mathcal{L}_X)h^{ij} = - 2 N K^{ij}\,,
\end{equation}
implies the relation
\begin{equation}\label{Dpsi}
D_v \psi^i{}_\mu = (D^i N) n_\mu - N \psi^l{}_\mu K^i{}_l\,.
\end{equation}
Next apply the identity (\ref{id1}) to $V_\mu = n_\mu$. The left-hand side, using (\ref{Dvn}), is given by
\begin{equation}
D_v (\psi^j{}_\mu K_{ji}) - D_i (\psi^j{}_\mu D_j N) = \psi^m{}_\mu(D_v K_{mi} - K_{li} K^l{}_m - D_i D_m N)\,,
\end{equation}
where we have on the right used (\ref{Dpsi}) and (\ref{Dphi}) in the form $D_i \psi^l{}_\mu = n_\mu K^l{}_i$. Thus, contracting with $\phi^\mu{}_{{, j}}$,
\begin{equation}
(\partial_\tau - \mathcal{L}_X) K_{ij} - K_{li} K^l{}_j - D_i D_j N = N R_{\rho\sigma\mu\nu} n^\rho \phi^\sigma{}_{{, i}} n^\nu \phi^\mu{}_{{, j}}\,.
\end{equation}
One can compare the above derivation with the standard derivation of $\frac{1}{N}$ times this equation (see e.g. Sect. 4.4.1 of the book
\cite{gourgoulhon20123+} or \cite{giulini2014dynamical}). The rest is linear algebra and is of course completely standard. Namely, on the
right-hand side one uses (\ref{compl}) to eliminate
$n^\mu$ by
\begin{equation}
n^\rho n^\nu = - g^{\rho\nu} + \phi^\rho{}_{{, l}}\phi^\nu{}_{,m}h^{lm}
\end{equation}
and the Gauss equation (\ref{Gauss}) to eliminate $R_{lijm} h^{lm}$. One finds
\begin{equation}
(\partial_\tau - \mathcal{L}_X) K_{ij}= N(R_{ij} - \mathcal{R}_{ij}) + 2 N K_{i\ell} K^\ell{}_j 
  - NK K_{ij} + D_i D_j N
\,.
 \label{13VIII25.5}
\end{equation}

\subsection{The Bianchi identities}

The equations governing the propagation of the scalar and vector constraints, i.e. 
\begin{equation}\label{C,Ci}
\mathcal{C} =: \frac{1}{2}(\mathcal{R} - 2 \Lambda   - K_{ij} K^{ij} + K^2) = 0\,,\hspace{1cm}\mathcal{C}_i := D_j(K_i{}^j - \delta_i{}^j K) = 0
 \,,
\end{equation}
 can be obtained directly from 
the explicit form of these constraints 
 by a brute force calculation, in which the vanishing or not of $N$ does not matter.
 It is however easier to derive them using their definition in terms of a slicing $(N,X^i)$ of a spacetime $(M,g_{\mu\nu})$,  again with $N$ possibly vanishing and/or changing sign, 
together with the contracted Bianchi identities for $g_{\mu\nu}$,
as follows.

Recalling that 
$$
 \mathcal{C}_i = \phi^\mu{}_{,i} n^\nu (G_{\mu\nu}+\Lambda  g_{\mu\nu}) = \phi^\mu{}_{,i} n^\nu G_{\mu\nu}
 \,,
 $$
from (\ref{CCia})-\eqref{CCib} we calculate as follows:
\begin{align}
D_v \mathcal{C}_i =  
& \
 D_i(N n^\mu) n^\nu G_{\mu\nu} + \phi^\mu{}_{,i} \phi^\nu{}_{,j} (D^j N) G_{\mu\nu} + N \phi^\mu{}_{,i} n^\nu n^\rho \nabla_\rho G_{\mu\nu} 
  \nonumber
 \\
=  & \ 
 (D_i N) (\mathcal{C} +\Lambda ) + N n^\mu \phi^\nu{}_{,j} K_i{}^j G_{\mu\nu} + \phi^\mu{}_{,i} \phi^\nu{}_{,j} (D^j N) G_{\mu\nu} 
  \nonumber
\\
 & \
   + N \phi^\mu{}_{,i}(- g^{\nu\rho} + h^{kl}\phi^\nu{}_{,k} \phi^\rho{}_{,l})\nabla_\rho  G_{\mu\nu}
  \nonumber
  \\
=  &\
(D_i N)(\mathcal{C} +\Lambda )
 \nonumber
 + N \phi^\mu{}_{,j} K_i{}^j G_{\mu\nu} + D^j(N \phi^\mu{}_{,i} \phi^\nu{}_{,j} G_{\mu\nu}) - N
  D^k(\phi^\nu{}_{,k} \phi^\mu{}_{,i})G_{\mu\nu}
 \\ 
= & \
  (D_i N)\mathcal{C}  + D^j[N(G_{ij}+\Lambda  h_{ij} )] - N K \mathcal{C}_i
   \nonumber
 \\ 
= & \
 (D_i N) \mathcal{C} + D^j[N (R_{ij} +
  (2\Lambda  + \mathcal{C} -h^{k\ell}R_{k\ell}) h_{ij})
   ] - N K \mathcal{C}_i\ 
  \,,
\end{align}
where we have used (\ref{Dvpsi})-(\ref{Dvn}) in the first line, the contracted Bianchi identity in the second line, a cancellation between the second term on the right and the expression $- N  \phi^\nu{}_{,k} (D^k \phi^\mu{}_{,i})G_{\mu\nu}$ coming from the last term in the third line, and the algebraic identity 
\begin{equation}
G_{ij} + \Lambda h_{ij}  = R_{ij} +  (2\Lambda  + \mathcal{C} -h^{k\ell}R_{k\ell}) h_{ij} 
\,,
\end{equation}
as well as $D^j \phi^\mu{}_{,j} = K n^\mu$,
 in the last line. Thus
\begin{equation}
 (\partial_\tau - \mathcal{L}_X) \mathcal{C}_i =
   D^j
   \big[N
   \big(
   R_{ij} 
    +(2\Lambda  - R^\ell{}_\ell) h_{ij}
     \big)
      \big] 
    + N D_i\mathcal{C} 
    + 2 (\partial_i N)\mathcal{C} - N K \mathcal{C}_i
\,.
 \label{3IV25.1}
\end{equation}
Similarly for $\mathcal{C} = n^\mu n^\nu (G_{\mu\nu} + \Lambda  g_{\mu\nu})$ we have
\begin{align}
D_v \mathcal{C}
  = & \
    2 (D^i N) \mathcal{C}_i + N n^\nu \phi^\mu{}_{,i} \phi^\rho{}_{,j} h^{ij} \nabla_\rho G_{\mu\nu} 
    \nonumber
 \\
    = & \
     2 (D^i N) \mathcal{C}_i + N D^i \mathcal{C}_i - N (D^i (\phi^\mu{}_{,i} n^\nu)) G_{\mu\nu}
      \,.
\end{align}
Thus
\begin{equation}
(\partial_\tau - X^l \partial_l)\mathcal{C} = N D^i \mathcal{C}_i + 2 (D^i N) \mathcal{C}_i - 2 N K \mathcal{C} 
 - N K^{ij}
 (
    R_{ij} 
    +(2\Lambda  - R^\ell{}_\ell) h_{ij} 
    ) 
 \,.
 \label{3IV25.2}
\end{equation}
Equations \eqref{3IV25.1}-\eqref{3IV25.2} become a closed set of symmetrisable-hyperbolic  equations for the constraints $(\cC,\cC_i)$,
first observed in \cite{Frittelli},
when e.g.
\begin{equation}\label{15VIII25.1}
 R_{ij} 
    +(2\Lambda  - R^\ell{}_\ell) h_{ij} =0
     \,,
\end{equation}
equivalently, when
$$
 R_{ij} 
     = \frac{2\Lambda}{n-1} h_{ij} 
     \,.
$$
As already pointed-out, we emphasise that more work is needed to show propagation of the scalar and vector constraints for solutions of the Anderson-York equations, because these solution  do  not directly provide a metric satisfying \eqref{15VIII25.1}.

\subsection{The Anderson-York equations}
 \label{s2III25.1}

The Anderson-York equations form a nonlinear first-order system of equations for the \emph{dynamical field} $\Phi$, defined as
\begin{equation}
 \label{4II25.52}
 \Phi:= (h_{ij}, K_{ij}, \chi_{ijk}) 
  \,,
\end{equation} 
where $h_{ij}$, $K_{ij}$ and  $ \chi_{ijk} $ are tensor fields  symmetric in the first two indices,  
where one  assumes that $\det h_{ij}$ has no zeros.
The equations are   
\begin{align}\label{4II25.21}
  &
(\partial_\tau - \mathcal{L}_X) h_{ij} =  2 N K_{ij} 
\,,
&
\\
 & 
(\partial_\tau - \mathcal{L}_X) K_{ij}=   \frac{N}{2} h^{kl}  
 \zD_l 
 \chi_{ijk} + 2 N K_{i\ell} K^\ell{}_j  - NK K_{ij}  
-\hat N_{ij}  
 \,,\label{4II25.21b}
\\ 
&
(\partial_\tau - \mathcal{L}_X) \chi_{ijk} =  2 N 
 \zD_k K_{ij}   + s_{ijk}
   \,,\label{4II25.21c}
 &
\end{align}
where $\zD$ denotes the covariant derivative-operator of  an arbitrarily chosen, possibly $\tau$-dependent, positive-definite tensor field $\zh_{ij}dy^i dy^j$, and
where the fields $s_{ijk}$ and $\hat N_{ij}$, to be defined in what follows, depend upon $\Phi$ but not its derivatives.
Inspection shows 
 that the system is symmetrisable-hyperbolic, with  the metric $h_{ij}$ satisfying an ODE along the integral curves of $\partial_\tau - X^\ell \partial_\ell$, and with a  scalar product for the pair of fields $(K_{ij},\chi_{ijk})$ that can be read from a  local $L^2$-energy equal to
\begin{equation}\label{5V25.1}
  \int  
  h^{ij}h^{k\ell} 
  \big( 
  K_{ik}K_{j\ell} +
   \frac 14 h^{pq}  \chi_{ikp}\chi_{j\ell q}
  \big)
  d\mu_h
  \,.
\end{equation}
One can add to it an expression such as, e.g., 
\begin{equation}\label{5V25.1ab}
  \int  
  \zh^{ij}\zh^{k\ell}  
  (h_{ik}-\zh_{ik})(h_{j\ell}-\zh_{j\ell}
  )
  d\mu_{\zh}
  \,,
\end{equation}
as a contribution to the energy from $h_{ij}$.
%

The causality properties of the system will be addressed in Section~\ref{s21I25.1}. 
\begin{center}
\fbox{
 \parbox{.829\textwidth}{\centering
In \eqref{4II25.21}-\eqref{4II25.21c} and unless explicitly indicated otherwise, $N$ is a shorthand for
 $Q\sqrt{\det h_{ij}}$, where $Q(\tau, y^i)$ is a prescribed density of weight $-1$ on each slice of constant $\tau$. }
  }
\end{center}
This redefinition of $N$ provides a crucial term in the equations to render them symmetric-hyperbolic.

In order to relate the equations above to the Einstein vacuum equations, we start by noting that \eqref{4II25.21} is simply the definition of the extrinsic curvature tensor $K_{ij}$  written as an evolution equation for $h_{ij}$ rather than the definition of $K_{ij}$.

Next, the derivatives of the metric will be encoded in a tensor field
$\chi_{ijk}$, in a somewhat roundabout way, after introducing
  a \emph{derivative constraint} tensor field $\cC_{ijk}$: 
\begin{equation}
\mathcal{C}_{ijk} := \chi_{ijk} - (\zD_{k} h_{ij} + 4 h_{k(i} \gamma_{j)})
 \,,
 \quad
  \mbox{ where } 
 \gamma_j =   h^{kl}\zD_{[j} h_{l]k}
  \,.
   \label{9III25.3}
\end{equation}
The relation \eqref{9III25.3} can be inverted to determine the derivatives of $h_{ij}$: 
\begin{equation}
\zD_{k} h_{ij} = \chi_{ijk} - \cC_{ijk}  + 4 h_{k(i} \mychi_{j)}\,, 
 \quad
  \mbox{ where } 
  \mychi_i  := \frac{1}{n-2}h^{jk}(\chi_{j[ki]}  - \cC_{j[ki]}  )
  \,.
   \label{9III25.12}
\end{equation}
The tensor field $\chi_{ijk}$ becomes related to the metric  when $\cC_{ijk}$ vanishes:
\begin{equation}
\chi_{ijk} = 
 \zD_{k} h_{ij} + 4 h_{k(i} \gamma_{j)}\,;
  \label{4II25.11b}
\end{equation}
equivalently, 
\begin{equation}
 \zD_{k} h_{ij} =  
  \chi_{ijk}  + 4 h_{k(i} \chi_{j)}
  \,, 
 \quad
  \mbox{ where } 
  \mychi_i   = \frac{1}{n-2}h^{jk}\chi_{j[ki]} 
  \,.
   \label{9III25.12cde}
\end{equation}

Let us denote by $\mathcal{R}_{ij}$ the Ricci tensor of the metric $h_{ij}$. Straightforward calculations 
lead to the formula
\begin{align}
\mathcal{R}_{ij} = 
  & \
   - \frac{1}{2} h^{kl}
   \big(\zD_k \zD_l h_{ij } + \zD_{(i}\zD_{j)} h_{kl } -  \zD_{j} \zD_kh_{ il}
    -  \zD_{i}  \zD_kh_{  j l}
    \big) + \bar r_{ij}(h,\zD h)
\,,
 \label{23III25.1}
\end{align}
where
$\bar r_{ij}$ depends upon the variables listed. 
We can rewrite this expression using the definition \eqref{9III25.3} of $\gamma_j$, i.e. 
\begin{equation} 
 \gamma_j =  \frac 12  h^{kl}
 \big(
  \zD_{j} h_{lk}
   -
   \zD_{l} h_{jk}
    \big) =  \frac 12  h^{kl}
 \big(
  \zD_{j} h_{lk}
   -
   \zD_{k} h_{jl}
   \big)
  \,.
   \label{9III25.3xd}
\end{equation}
This gives
\begin{align}
\mathcal{R}_{ij}   - \frac{1}{2} h^{kl}
   \zD_{(i}\zD_{j)} h_{kl } =  
  & \
  - \frac{1}{2} h^{kl}
    \zD_k \zD_l h_{ij } - \zD_i  \gamma_j  - \zD_j \gamma_i 
   +
    \bar r_{ij}   (h,\zD h )
    \nonumber
\\ =  
  & \
  - \frac{1}{2} \zD^l
   \big(  \zD_l h_{ij } +4  h_{l(i} \gamma_{j)}
   \big) + \hat r_{ij}   (h,\zD h )    
\,,
 \label{23III25.1nb}
\end{align}
with a field $\hat r_{ij}$ defined by the above sequence of equalities. 

The (explicit) second-derivative terms at the left-hand side will be handled by replacing $N$ with $Q$. For this we start with the replacement
\begin{equation}\label{4II25.2d}
  N = \sqrt{\det h}\, Q
\end{equation}
which gives
\begin{align}
 D_i D_j  N
  = & \
     \sqrt{\det h} 
 \Big(
  D_i D_j  Q + \frac  Q2 h^{k\ell} \zD_{(i}\zD_{j)} h_{k\ell} + \breve N_{ij}(Q,D Q,h,\zD h)
  \Big)
 \nonumber
\\
  = & \
   \underbrace{\frac  N2 h^{k\ell} \zD_{(i}\zD_{j)} h_{k\ell} }_{(\diamond)}
    + \bar N_{ij}(Q,\zD Q,\zD^2 Q,h,\zD h) 
    \,,
\end{align}
with functions $\breve N$ and $\bar N$ defined by the last two equalities; the point is, that these functions again do not involve second derivatives of $h$. This allows us to rewrite \eqref{23III25.1nb} as
\begin{eqnarray}
 N \mathcal{R}_{ij} -
   D_i D_j N  
  & = &
     - \frac{N}{2} h^{kl} \zD_k
   \big(  \zD_l h_{ij } +4  h_{l(i} \gamma_{j)}
   \big) + \bar r_{ij}     + \bar r_{ij}   (h,\zD h )
   -\bar N_{ij}  
  \nonumber 
\\
   & =   & 
    \frac{N}{2} h^{kl} \zD_k
 \big( \cC_{ijl} -  \chi_{ijl}
 \big)
   - \check N _{ij}(h,\chi,\cC)
\,,
\label{23III25.2}
\end{eqnarray}
with $\check N_{ij}$  obtained from   $\bar N_{ij}  -\bar r_{ij}  $ by rewriting every occurrence of $\zD h$ there 
by 
$\chi_{ijk}-\cC_{ijk}$ using \eqref{9III25.12}.  

In what follows we  always suppose that 
the tensor field $h_{ij}dx^idx^j$ is positive-definite throughout the region under consideration.  

As shown in Section~\ref{ss9III25.1} (see \eqref{13VIII25.5}),
for any slicing it holds that
\begin{equation}
 NR_{ij}  = (\partial_\tau - \mathcal{L}_X) K_{ij}
 +  N \mathcal{R}_{ij} - D_i D_j N
 - 2 N K_{i\ell} K^\ell{}_j  +  NK K_{ij} 
  \,,
  \label{4II25.3}
\end{equation}
where $R_{ij}$ is the space-projection  of $R_{\mu\nu}$ (regardless of whether or not $N$ has zeros). 
Using \eqref{23III25.2}, Equation~\eqref{4II25.3} can be rewritten as
\begin{equation}
(\partial_\tau - \mathcal{L}_X) K_{ij}=   
 \frac{N}{2} h^{kl}  \zD_l 
 (\chi_{ijk} - \cC_{ijk})  - N R_{ij} 
 - \check N_{ij}    + 2 N K_{i\ell} K^\ell{}_j  - NK K_{ij}  
 \,.
  \label{4II25.4- }
\end{equation}
The equation
\begin{equation}
(\partial_\tau - \mathcal{L}_X) K_{ij}=   \frac{N}{2} h^{kl}  
 \zD_l 
 \chi_{ijk}  
-\hat N_{ij}
   + 2 N K_{i\ell} K^\ell{}_j  - NK K_{ij}  
 \,.
  \label{4II25.4}
\end{equation}
is obtained from \eqref{4II25.4- } by setting  $\cC_{ijk}\equiv 0$ 
and  imposing the space-part of the vacuum Einstein equations,
$$
 R_{ij}=\lambda  h_{ij}
 \,.
$$
This  defines the field 
\begin{equation}\label{25IV25.2}
 \hat N_{ij} :=  
  \check  N_{ij} |_{\cC_{ijk}=0} 
  + \lambda N h_{ij}
 \end{equation}
  appearing in the evolution equation \eqref{4II25.21b}.


For further purposes we  note that when \eqref{4II25.4} is satisfied, 
we can insert it back  into \eqref{4II25.3} to obtain, after using \eqref{23III25.2}
and \eqref{25IV25.2},  
\begin{align}
 N (R_{ij} - \lambda h_{ij})
    =  
    \frac{N}{2} h^{kl}  
     \zD_l  \mathcal{C}_{ijk} 
    + \check  N_{ij} |_{\cC_{ijk}=0} 
    - \check  N_{ij}  
  \,.
  \label{4II25.3b}
\end{align}

It remains  to elucidate the origin of \eqref{4II25.21c}. For this let us return to \eqref{4II25.21}, i.e.\
\begin{equation}\label{Dvha}
(\partial_\tau - \mathcal{L}_X) h_{ij} =  2 N K_{ij}
 \,.
\end{equation}
Having in mind \eqref{9III25.12cde}, we will obtain an equation for the field  $\chi_{ijk} $ 
by differentiating \eqref{Dvha} and commuting derivatives. For this we start with an equation for the field $\gamma_i$ of \eqref{9III25.3xd}, which is readily derived to be
\begin{equation}
(\partial_\tau - \mathcal{L}_X) \gamma_i =  - N \mathcal{C}_i 
 + s_i
  \,,
\end{equation}
with $\cC_i$ given by \eqref{C,Ci},
where 
$$
 s_i\equiv s_i(K_{ij},h_{ij},\chi_{ijk}-\cC_{ijk})
$$
depends on the   dynamical fields $\Phi$ and on $\cC_{ijk}$,
but not on their derivatives. 
Subsequently one obtains  
\begin{equation}\label{Dvchiagain}
(\partial_\tau - \mathcal{L}_X)
\big(\mathring{\nozD}_{k} h_{ij} + 4 h_{k(i} \gamma_{j)}
\big) =  2 N (\zD_k K_{ij} - 2\,h_{k(i}\mathcal{C}_{j)}) + \breve s_{ijk}
 \,,
\end{equation} 
where the field, as defined by this equation, 
$$
 \breve s_{ijk}=
  \breve s_{ijk}(K_{ij},h_{ij},\chi_{ijk}-\cC_{ijk})
$$
likewise
depends on the   dynamical fields $\Phi$ and on $\cC_{ijk}$,
but not on their derivatives. 
This can be rewritten as  
\begin{equation}\label{Dvchiolala}
(\partial_\tau - \mathcal{L}_X) \chi_{ijk} =  2 N (\zD_k K_{ij} - 2\,h_{k(i}\mathcal{C}_{j)}) + 
 \breve s_{ijk} +(\partial_\tau - \mathcal{L}_X) \cC_{ijk} 
 \,.
\end{equation} 
Defining
$$
s_{ijk} = \breve s_{ijk}|_{\cC_{ijk}=0}
$$
 and setting to zero the remaining occurrences of  $\cC_{j}$ and $\cC_{ijk}$ in  \eqref{Dvchiolala}, one obtains
\begin{equation}\label{Dvchia}
(\partial_\tau - \mathcal{L}_X) \chi_{ijk} =  2 N \zD_k K_{ij}   + s_{ijk} 
 \,,
\end{equation} 
which is \eqref{4II25.21c}. 
Note that to arrive at this equation we invoked 
the vacuum vector constraint equation, without which the symmetrisation would not have occurred. 
 
To continue, suppose that we have a solution of  \eqref{4II25.21}-\eqref{4II25.21c}, in particular of \eqref{Dvchia}.  
Inserting this last equation back into  \eqref{Dvchiolala} we obtain 
\begin{equation}\label{Dvchixc}
(\partial_\tau - \mathcal{L}_X) \cC_{ijk}  =  
 4 N  h_{k(i}\mathcal{C}_{j)}
  + 
\breve s_{ijk}|_{\cC_{ijk}=0} - \breve s_{ijk}  
 \,.
\end{equation} 
%
Next, we can calculate the associated constraint functions $(\mcC,\mcC_i)$ defined in
\eqref{C,Ci}.  (In particular $\mathcal{C}$ is of second differential order in $h_{ij}$ and $\mathcal{C}_i$ is of first differential order in $h_{ij}$.) Using \eqref{3IV25.1},  \eqref{3IV25.2}, \eqref{4II25.21}  
 and \eqref{4II25.3b} we obtain the following further equations satisfied by the fields $ (\mathcal{C}, \mathcal{C}_i, \mathcal{C}_{ijk}) $: 

%
\begin{align}\label{n4II25.25} 
(\partial_\tau - \mathcal{L}_X)
 \cC  = \
&    -  K^{ij}
\Big[
 \red{\frac1 2} N
  \zD^k \big( \mathcal{C}_{ijk} 
  - h_{ij} \mathcal{C}^\ell{}_{\ell k}\big)  
  \nonumber
\\
 &    
    + \check  N_{ij} |_{\cC_{ijk}=0} 
    - \check  N_{ij} 
     - h_{ij} h^{k\ell} \big( \check  N_{k\ell} |_{\cC_{ijk}=0} 
    - \check  N_{k\ell} 
    \big) 
     \Big]
  \nonumber
\\
  &
  + N D^i \mathcal{C}_i - 2 NK \mathcal{C} + 2 (D^i N) \mathcal{C}_i 
    \,, 
\\ 
 (\partial_\tau - \mathcal{L}_X)\, \mathcal{C}_i =
 \ 
 &  
    D^j\Big( \red{\frac 12}N \zD^k  \big(\mathcal{C}_{ijk} 
  - h_{ij} \mathcal{C}^\ell{}_{\ell k} 
  \big)
  \nonumber
\\
 &    
    + \check  N_{ij} |_{\cC_{ijk}=0} 
    - \check  N_{ij} 
     - h_{ij} h^{k\ell} \big( \check  N_{k\ell} |_{\cC_{ijk}=0} 
    - \check  N_{k\ell} 
    \big) 
  \Big) 
  \nonumber
  \\
  &
  + N \partial_i \mathcal{C} + 2 (\partial_i N) \mathcal{C} - N K \mathcal{C}_i 
   \,. 
    \label{8III25.2End}
\end{align}

Suppose, now,  that all the constraints are satisfied at $\tau=0$. Recall that the vanishing of $\cC$ and $\cC_i$  is a standard necessary condition to construct a vacuum spacetime out of the Cauchy data. For $\cC_{ijk}$ we simply define $\chi_{ijk}|_{\tau=0}$ so that $\cC_{ijk}$ vanishes at $\tau=0$, hence the vanishing of the initial values of $\cC_{ijk}$ does not provide any restrictions on $(h_{ij},K_{ij})|_{\tau=0}$. 

Commuting the equations with $\partial_\tau$ and with $D_i$ one finds by induction that  $\tau$- and space-derivatives of any order of $(\cC,\cC_i,\cC_{ijk})$ vanish at $\tau=0$. 


\subsection{The evolution of the constraints}
 \label{s18VIII25.1}

We are ready now to prove Theorem \ref{T1VI25.1}.  
%
We start by rewriting the equations as a first-order system. For this we introduce 
\begin{equation}
\mathcal{D}_{ijkl} :=\mathring{D}_{l} \mathcal{C}_{ijk}
 \,,
 \quad
 \mathcal{D}_{ij} := \mathcal{D}_{ijs}{}^s
\,.
\end{equation}
We have 
\begin{equation}
 \label{31VII25.1}
(\partial_\tau - \mathcal{L}_X)\mathcal{D}_{ijkl} =  4 N h_{(i|k} \mathring{D}_{l|} \mathcal{C}_{j)} + O(\mathcal{C}_i) + O(\mathcal{C}_{ijk})+ O(\mathcal{D}_{ijkl})
\,,
\end{equation}
\begin{equation}
 \label{31VII25.3a}
(\partial_\tau - \mathcal{L}_X)\mathcal{C} = N D^i\mathcal{C}_i + O(\mathcal{C}_i) + O(\mathcal{C}) + O(\mathcal{D}_{ijkl})
\,,
\end{equation}
\begin{equation}
 \label{31VII25.3}
(\partial_\tau - \mathcal{L}_X)\mathcal{C}_i = 
\frac{N}2(D^j\mathcal{D}_{ijm}{}^m-D_i \mathcal{D}_m{}^m{}_s{}^s) + N D_i \mathcal{C} +
O(\mathcal{C}_i) + O(\mathcal{C}) + O(\mathcal{D}_{ijkl})
\,.
\end{equation} 

We will show that this system,  adjoined with \eqref{Dvchixc},  is 
 \emph{microlocally symmetrisable hyperbolic} as in Taylor~\cite[Equations~(5.2.26)-(5.2.27)]{Taylor91}. This notion is defined there
as  symmetrisability in Fourier space of the principal symbol,  
with a positive-definite symmetriser smooth in all its arguments and homogeneous of degree zero in the Fourier variables, 
and referred to there simply as \emph{symmetrisable hyperbolic}, but the latter terminology is  used here (and in several standard references)%
\footnote{Taylor's definition is very similar to that of \emph{strong hyperbolicity} in some, but not all, references.}
to denote instead joint symmetrisability of the individual matrices appearing in the principal part of the equation. To avoid ambiguities we use distinct terminologies. 

Indeed, we can write the system as
\begin{equation}\label{5VIII25.1}
  \partial_\tau f  = \mathcal{U}^i D_i f + \mathcal{W}
  \,,
\end{equation} 
where $\mathcal{W}$ does not contain any derivatives of $f$. For  $k_i\in T^*\hyp$  we then need to analyse the principal symbol
$$
 \sigma(k):= \mathcal{U}^i k_i
 \,.
$$

\subsubsection{Diagonalisability of the principal symbol}
 \label{ss5VIII25.1}

Letting $\omega\in\R$ denote an eigenvalue, the eigenvectors for the principal symbol of the space-part of the system \eqref{Dvchixc}, \eqref{31VII25.1}-\eqref{31VII25.3}  are solutions of the equations
\begin{equation}\label{inspec1}
(\omega - X^l k_l) \mathcal{C} = N k^l \mathcal{C}_l
 \,,
\end{equation}
\begin{equation}\label{inspec2}
(\omega - X^l k_l) \mathcal{C}_i = \frac{N}{2}(k^j \mathcal{D}_{ij} - k_i \mathcal{D}_r{}^r) + N k_i \mathcal{C} 
 \,,
\end{equation}
\begin{equation}\label{inspec4}
(\omega - X^l k_l) \mathcal{C}_{ijk} = 
 0
 \,,
\end{equation}
\begin{equation}\label{inspec3}
(\omega - X^r k_r) \mathcal{D}_{ijkl} = 4 N h_{(i|k} k_{l|} \mathcal{C}_{j)}
 \,.
\end{equation}
%

We start with the case  $N=0$, where it can be easily seen that
\begin{equation}\label{5VIII25.11}
\omega = \omega_0 := X^\ell k_\ell  
\end{equation}
is an eigenvector and every vector is an eigenvector.

In fact, $\omega_0$
is an eigenvalue independently of whether or not $N$ vanishes. Indeed, when $N\neq 0$, one checks that the associated eigenvectors $f_0$ in $\mathbb{R}^{K}$, where
\begin{equation}\label{8VIII25.12}
 K:=
  n+1+\frac{n^2(n+1)}{2}+ \frac{n^3(n+1)}{2}  
 \,,
\end{equation}
take the form
 \wancheck{5VIII}
\begin{displaymath}
f_0 =
\left(\begin{array}{c}
\mathcal{C}_{ijl} = \mathcal{C}^0_{ijl}\\
\mathcal{C} = \mathcal{C}^0\\
\mathcal{C}_i = 0\\
\mathcal{D}_{ijlm} = \mathcal{E}^0_{ijlm} + \frac{2}{n(n-1)} \mathcal{C}^0 q_{ij} h_{lm}
\end{array} \right)
\,,
\end{displaymath}
with arbitrary $\mathcal{C}^0_{ijl}$, $\mathcal{C}^0$, and $\mathcal{E}^0_{ijlm}$ subject to
\begin{equation}
\mathcal{E}^0_{ijlm} k^i q^j{}_{j'} h^{lm} = 0\,,\hspace{1cm}\mathcal{E}^0_{ijlm} q^{ij} h^{lm} = 0
 \,,
\end{equation}
where  $q_{ij} = h_{ij} - \frac{k_i k_j}{|k|_h^2}$, with $|k|_h^2 = h^{ij} k_i k_j$.
Here and below, the coefficients with superscripts 0,1,2 refer to free parameters. Thus $f_0$ has $K - 2 n$ free parameters.

Next, 
\begin{equation}\label{5VIII25.12}
\omega = \omega_\pm := X^l k_l \pm N |k|_h
\end{equation}
provides two further eigenvalues.  The corresponding $2n$-parameter set of eigenvectors $f_\pm$ is described by
\wancheck{5VIII}
\begin{displaymath}
f_\pm =
\left(\begin{array}{c}
\mathcal{C}_{ijl} = 0\\
\mathcal{C} = \pm k^l \bar{\mathcal{C}}_l/|k|_h\\
\mathcal{C}_i = \bar{\mathcal{C}}_i \\
\mathcal{D}_{ijlm} = \pm 4 h_{(i|l} k_{m|} \bar{\mathcal{C}}_{j)}/|k|_h 
\end{array} \right)
\,.
\end{displaymath}
To see that the $f_\pm$'s are eigenvectors, note that (\ref{inspec1})-(\ref{inspec3}) imply that 
$$
 [(\omega - X^l k_l)^2 - N^2 |k|_h^2]\mathcal{C}_i = 0
  \,.
$$
The remaining components are also easy to check.

It remains to check that the $K$ vectors comprised by $f_0, f_\pm$ span the whole space. For this it is convenient to define
\begin{displaymath}
f_1 \equiv \frac{f_+ + f_-}{2} := 
\left(\begin{array}{c}
\mathcal{C}_{ijl} = 0\\
\mathcal{C} = 0\\
\mathcal{C}_i = \mathcal{C}^1_i \\
\mathcal{D}_{ijlm} = 0
\end{array} \right)
 \,,
\end{displaymath}
\begin{displaymath}
f_2 \equiv  \frac{f_+ - f_-}{2} 
 :=
\left(\begin{array}{c}
\mathcal{C}_{ijl} = 0\\
{\mathcal{C}} = k^l \mathcal{C}^2_l /|k|_h\\
\mathcal{C}_i = 0\\
\mathcal{D}_{ijlm} = 4 h_{(i|l} k_{m|} \mathcal{C}^2_{j)}/|k|_h 
\end{array} \right)
 \,.
\end{displaymath}
It is now straightforward that the equation
$f_0 + f_1 + f_2 = 0$ implies $\mathcal{C}^0_{ijl} = \mathcal{C}^0 = \mathcal{E}^0_{ijlm} = \mathcal{C}^1_i = \mathcal{C}^2_i = 0$. 
\wancheck{6VIII}

We conclude that, for $k_i\ne 0$, the symbol $\sigma(k)$ has a full set of real eigenvectors.

\subsubsection{Connecting with Theorem 5.2.D in Taylor~\protect\cite{Taylor91}}
\label{ss16VIII25.1}
 
Consider a first-order system of equations of the form
\begin{equation}
\partial_\tau f^A = \mathcal{U}^{(l)A}{}_B (\tau,y,f) \partial_l f^B + \mathcal{W}^A(\tau,y,f)\,,\hspace{1cm}A,B = 1,....K
 \,.
 \label{16VIII25.0}
\end{equation}
where the fields $f^A$ are sections of a bundle, say $\bundle$, over $\R\times \hyp$, with fibers of dimension $K$. In local coordinates, 
for every $(\tau,y,k)\in \R\times \hyp\times T^*_y\hyp$ the principal symbol $\sigma(k)$ of the differential operator at the right-hand side defines a map 
of 
$\mathbb{R}^K$ into itself,  which we write as
$$\mysigma^A{}_B
\equiv \mysigma(k)^A{}_B :=\mathcal{U}^{(l)A}{}_B 
k_l
 \,.
$$
For each $ 0 \ne  k \in  T^* \hyp$ let  $G(k)$ be  a (positive-definite) scalar product on fibers of $\bundle$. The tensor field $G(k)$ is called a \emph{microlocal symmetriser} for the system \eqref{16VIII25.0} if $\sigma(k)$ is symmetric with respect to $G(k)$, i.e.\ if
for all $X,Y$ tangent to the fibers of $\bundle$ and for all $k\in T^*\hyp$ we have
\begin{equation}\label{16VIII25.1}
  G(k)(\sigma(k)X,Y) = G(k)(X,\sigma(k) Y)
  \,. 
\end{equation}
The field $G(k)$ is called a \emph{symmetriser} if $G(k)$ is independent of $k$.

Writing   $G(k)$ in local coordinates as 
$G(k)_{AB} \equiv G(k)_{AB} (\tau,y,f;k)$,
the condition \eqref{16VIII25.1} is equivalent to
%
\begin{equation}\label{16VIII25.2a}
  G(k)_{AB} \sigma(k)^A{}_{A'}    = G(k)_{A'B'}   \sigma(k)^{B'}{}_{B}  
  \,, 
\end{equation}
or, since $ G(k)_{A'B'}=G(k)_{B'A'}$,
%
\begin{equation}\label{16VIII25.3a}
  G(k)_{BA} \sigma(k)^A{}_{A'}    = G(k)_{A'A}   \sigma(k)^{A}{}_{B}  
  \,. 
\end{equation}
In plain English: the tensor field $G(k)\sigma(k)$ is a symmetric two-covariant  tensor on the fibers of $F$.

It is proved by Taylor in  \cite[Theorem 5.2.D]{Taylor91} that a quasilinear system of the form \eqref{16VIII25.0}
is well-posed in suitable Sobolev spaces provided that a microlocal symmetriser $G(k)  $ for $\mysigma (k)  $ exists, with $G(k)$ positive definite, and    homogenous of degree 0 in $k  \in \mathbb{R}^n \setminus \{0\}$, and   smooth in all its arguments  away from $\{k =0\}$. 

Note that the standard definition of symmetrisable-hyperbolic is existence of a symmetriser which is independent of $k $, which is the case for e.g.
 the Anderson-York equations.

Now, the existence of a complete set of real eigenvectors clearly implies the existence of a microlocal symmetriser: Indeed, at every point of the fiber of $F$ we can choose a diagonalising basis for $\sigma(k)$, and take $G(k)$ to be a scalar product for which this basis is orthogonal. However, this procedure does not warrant that $G(k)$ will have the right continuity and/or smoothness properties.
In order to address this last issue, recall that existence of a diagonalising basis for $\sigma(k)$ implies the existence of an invertible matrix $S^A{}_B\equiv S^A{}_B(k)$ such that the tensor field
\begin{equation}\label{symdelta}
\hat \mysigma_{AB}:=  \delta_{EA} S^E{}_{C} \mysigma^C{}_{D} S^{- 1\,D}{}_{B}
\end{equation}
is diagonal, in particular symmetric in $(A,B)$. Here and below, all of $
\delta_{AB}$,  $
\delta^{AB}$, and  $
\delta_{A}^{B}$ denote the Kronecker symbol.

We note that
\begin{align}\label{symdelta2}
 S^{- 1\,C'}{}_{A'} \delta^{A'A}\hat \mysigma_{AB} S^{B}{}_{D'}
 = \ &
 S^{- 1\,C'}{}_{A'} 
  \underbrace{
    \delta^{A'A}\delta_{EA} S^E{}_{C}
    }_{S^{A'}{}_C} \mysigma^C{}_{D}
  \underbrace{
   S^{- 1\,D}{}_{B} S^{B}{}_{D'}}_{
   \delta^D_{D'}}
  \nonumber
\\  
 = \ &  
  \mysigma^{C'}{}_{D'}   
  \,.
\end{align}
We set
%
\begin{equation}
G(k)_{DC'} = S^{A'}{}_D S^{C}{}_{C'}  \,\delta_{A'C}
 \,.
 \label{16VIII25.8}
\end{equation}
Then
\begin{align}\label{16VIII25.6}
G(k)_{DC'} 
      \underbrace{
       \mysigma(k)^{C'}{}_{D'}
       }_{ S^{- 1\,C'}{}_{B'} \delta^{B'A}\hat \mysigma_{AB} S^{B}{}_{D'}
       } 
     \nonumber
    = \
     &      
     S^{A'}{}_D 
      \underbrace{
      S^{C}{}_{C'}  
       \delta_{A'C }
       S^{- 1\,C'}{}_{B'}}_{\delta^C_{B'} 
       \delta_{A'C }= \delta_{A'B'}}
         \delta^{B'A}\hat \mysigma_{AB} S^{B}{}_{D'}
     \nonumber
\\ 
    = \
     &    
     S^{A'}{}_D  
       \underbrace{
        \delta_{A'B' } 
         \delta^{B'A}
         }_{\delta^A_{A'}
         }
         \hat \mysigma_{AB} S^{B}{}_{D'}
     \nonumber
\\ 
    = \
     &    
     S^{A }{}_D   \hat \mysigma_{AB} S^{B}{}_{D'}
     \,.
\end{align}
Since $\mysigma_{AB}$ is symmetric, the last expression does not change under exchange of $D$ with $D'$. Hence
\begin{align}\label{16VIII25.7} 
G(k)_{DC'} 
       \mysigma(k)^{C'}{}_{D'} 
    = \
     &    
G(k)_{D'C'} 
       \mysigma(k)^{C'}{}_{D} 
     \,,
\end{align}
as needed 
for a microlocal symmetriser for $\sigma(k)$.

The tensor field $G(k)$ is positive definite on the fibers of $F$: for $X\ne 0$ and $k
\ne 0$ we have, using hopefully obvious notation,
$$
 G(k)(X,X) = 
 |S(X)|^2_\delta >0 
\,,
$$
since $S$ is invertible. 

Finally, $G(k)$ will have the right smoothness and homogeneity properties if $S(k)$ does.

In order to apply this to the system of equations governing the propagation of constraints,  let $T(k)^A{}_B$ be the column matrix formed by the eigenvectors of Section~\ref{ss5VIII25.1}.
This matrix is smooth in $(\tau,y;k)$ away from the set $k=0$. We denote by $S(k)^A{}_B$   its inverse. Both fields are homogenous in $k $ of degree zero, and $S^A{}_B$
satisfies (\ref{symdelta}).  
 We conclude that the theorem by Taylor applies, and provides
  existence and uniqueness of solutions for sufficiently regular Sobolev-class initial data sets on $\R^n$. 
 
 One can now pass to general initial data sets on general manifolds using the causality properties of the equations governing the evolution of the constraints, as analysed in Section~\ref{ss8VIII25.2} below, and standard arguments.

%
\section{Causality}
 \label{s21I25.1}

We determine the space metric  $h_{ij}$ 
using the evolution equations \eqref{4II25.21}-\eqref{4II25.21c}. 
In order to pass from initial data on  $\R^n$, with controlled asymptotics, to general initial data on a general manifold, one needs to control the domains of dependence and influence associated with these equations.
Even more importantly, controlling the geometry of the domains of dependence for the system of equations satisfied by the constraints is the key for obtaining solutions of the Einstein equations. Indeed, faster-than-light, respectively infinite, propagation speed for the constraint equations would result in smaller, respectively empty, regions where the metric will be Einstein, except perhaps for  restricted classes of initial data which would require a case-by-case analysis. 

 In regions where $|N|>0$ one expects the domain of dependence to be the usual one, as defined by the spacetime metric, but the question arises what happens when zeros of $N$ occur, or when $N$ changes sign. This can be answered by using the results in~\cite{JMR,Rauch1,Rauch2}, compare~\cite{MetivierSym}.
Our terminology below will be a mixture of the terminology in these papers and of the usual terminology of general relativistic causality theory as, e.g., in Chapter~2 of \cite{ChBlackHoles}. 

\subsection{Anderson-York equations}
 \label{ss8VIII25.1}
 
In order to address the problem at hand for  the Anderson-York evolution equations, in these equations  we replace 
$(\partial_\tau,\partial_i)$  by   
$$
 (\myxitau,\myxi_i)\equiv (\myxitau,\myxi) \equiv (\myxi_\mu)
 \,.
$$
After this substitution, the principal 
symbol 
of the equations takes the form
\begin{equation}\label{21I25.1}
             \left(
               \begin{array}{c}
                 (\myxitau - X^l\myxi_l) h_{ij}  \\
                 (\myxitau - X^l\myxi_l) K_{ij} -  \frac{N}{2} h^{kl} \myxi_l\chi_{ijk} \\
                 (\myxitau - X^l\myxi_l) \chi_{ijk} - 2N\myxi_k K_{ij} \\
               \end{array}
             \right)
             \,.
\end{equation}
The $h_{ij}$-part of the principal part of the equations decouples, and the $(ij)$-indices do not mix in \eqref{21I25.1}. So, to understand the propagation properties of this system it suffices to consider the following linear map involving a scalar field $\kappa$ and a covector field $\chi_k$:
\begin{equation}\label{21I25.2}
             \left(
               \begin{array}{c} 
                 (\myxitau - X^l\myxi_l) \kappa -  \frac{N}{2} h^{kl} \myxi_l\chi_{k} \\
                 (\myxitau - X^l\myxi_l) \chi_{k} - 2N\myxi_k \kappa \\
               \end{array}
             \right)
             =: 
  \sigma(\myxitau,\myxi)\left(
               \begin{array}{c}
                 \kappa \\
                 \chi_k \\
               \end{array}
             \right)
             \,.
\end{equation}
According to \cite{JMR}, the first step to determine the causality properties of a system with principal part encoded by \eqref{21I25.2} is to calculate the determinant of the linear map so defined. In order to do this, given $p\in \R\times \hyp$ and $\myxi \in T^*_pM$ we can choose a coordinate system in which $h_{ij}=\mathrm{diag}(1,\ldots,1)$ and 
$$
 \myxi =(\myxi_1,0,\ldots,0)=
 (|\myxi|_h,0,\ldots,0)
  \,,
$$
where $|\myxi|_h$ is the length of $\myxi$ in the metric $h$. 
Using the notation
$$
 \mychi=(\mychi_1,\mychi_\perp)
 \,,
$$
the map $ \sigma(\myxitau,\myxi)$ of  \eqref{21I25.2} can be rewritten as
\begin{eqnarray}\label{21I25.3}
  \sigma(\myxitau,\myxi)\left(
               \begin{array}{c} 
                 \kappa \\
                 \chi_1 \\
                 \chi_\perp
               \end{array}
             \right)
              & = &
 \left(
               \begin{array}{c} 
                 (\myxitau - X^1\myxi_1) \kappa -  \frac{N}{2}   \myxi_1\chi_{1} \\
                 (\myxitau - X^1\myxi_1) \chi_{k} - 2N\myxi_1  \delta_k^1 \kappa \\
               \end{array}
             \right)
             \nonumber
\\             
              &=
             & \left(
                \begin{array}{ccc}
                  \myxitau - X^1\myxi_1  & -  \frac{N}{2}   \myxi_1 & 0 \\
                 -  2N   \myxi_1 &  \myxitau - X^1\myxi_1  & 0 \\
                  0 & 0 & \myxitau - X^1\myxi_1  \\
                \end{array}
              \right)
              \left(
                \begin{array}{c}
                  \kappa \\
                  \chi_1 \\
                  \chi_\perp \\
                \end{array}
              \right)
             \,.
             \phantom{xxxxx}
\end{eqnarray}
The calculation of the associated \emph{characteristic polynomial} $p(\myxitau,\myxi)$, defined as the determinant of the matrix $\sigma(\myxitau,\myxi)$, is   straightforward
\begin{eqnarray}\label{22I25.4}
  p(\myxitau,\myxi) 
  &
  =
  & (\myxitau - X^1\myxi_1)^{n-1}
  \big(
  (\myxitau - X^1\myxi_1)^2 - N^2 (\myxi_1)^2
   \big)
    \,.
\end{eqnarray}  
It follows that in a general coordinate system we will have 
\begin{eqnarray}\label{22I25.5}
  p(\myxitau,\myxi) 
  &
  =
  & (\myxitau - X^\ell \myxi_\ell)^{n-1}
  \big(
  (\myxitau - X^\ell\myxi_\ell)^2 - N^2 |\myxi|_h^2
   \big)
    \,.
\end{eqnarray}  

Taking into account the $h_{ij}$-part of the principal symbol of the system of main interest only changes the power of the $(\myxitau - X^\ell \myxi_\ell)$-factor in \eqref{22I25.5}, which does not affect the analysis below. Likewise the fact that the scalar $\kappa$ has to be replaced by $K_{ij}$, etc., results in replacing   $p(\myxitau,\myxi) $ by a power thereof, without affecting what follows.

One defines
\begin{equation}\label{22I25.6a}
  \myxitau^{\max{}}:= \max\{\myxitau: p(\myxitau,\myxi)=0\}
  \,.
\end{equation}
From \eqref{22I25.5} one finds
\begin{equation}\label{22I25.6b}
  \myxitau^{\max{}}=  X^\ell\myxi_\ell + |N| |\myxi|_h 
  \,.
\end{equation}
The   \emph{cone of  time-oriented timelike covectors} $\mcTpp$ at $p\in M$ is defined as  
\begin{eqnarray} 
  \mcTpp & :=& \{ (\myxitau,\myxi)\in T^*_pM: \myxitau > \myxitau^{\max{}} \} 
   \nonumber
\\
    &=&  \{ (\myxitau,\myxi)\in T^*_pM: \myxitau >    X^\ell\myxi_\ell + |N| |\myxi|_h \} 
    \,.
\end{eqnarray}
We see that $\mcTpp$ is that connected component of the quadratic cone 
\begin{eqnarray}
  \mcTp  &: =&  
   \{ (\myxitau,\myxi)\in T^*_pM: (\myxitau  -  X^\ell\myxi_\ell)^2 > N^2 h^{ij}\myxi_i\myxi_j \} 
   \,,
\end{eqnarray}
on which $\myxitau-X^\ell\myxi_\ell>0$.

In the terminology of~\cite{JMR}, 
as adapted to the manifold setting here, the \emph{propagation cone} $\Gammapp$ at $p$ is defined as 
\begin{eqnarray}
  \Gammapp & :=& \{ (Y^\tau,Y)\in T_p M: 
  \forall \ (\myxitau,\myxi)\in \mcTpp \quad
  Y^\tau \myxitau+ Y^i  \myxi_i\ge 0 \} 
    \,.
\end{eqnarray}

There arise two cases:

\begin{enumerate}
  \item Suppose that $N$ vanishes at $p\in M$. Then 
\begin{eqnarray} 
  \mcTpp 
    &=&  \{ (\myxitau,\myxi)\in T^*_pM: \myxitau >    X^\ell\myxi_\ell   \} 
    \,,
\end{eqnarray}
which is an open half-space bounded by the hyperplane  
$$ \mcNp:=\{(\myxitau,\myxi)\in T^*_pM: \myxitau =    X^\ell\myxi_\ell \} = \partial \mcTpp
\,.
$$
Let
$$
  \theta^0=d\tau\,,
  \quad
  \theta^i= dx^i + X^i d\tau
  \,,
$$
thus $\{\theta^\mu\}_{\mu=0}^n$ is a basis of $T^*_pM $ such that $\theta^1,\ldots,\theta^n$ spans the hyperplane $\mcNp $ and $\theta^0\in \mcTpp$. Decomposing covectors $\alpha \in T_p^*M$ in this basis as $\alpha = \alpha_\mu\theta^\mu$, $\alpha$ will belong to $\mcTpp$ if and only if $\alpha_0>0$.  Let 
$$
  e_0=\partial_\tau -X^i\partial_i
  \,,
  \quad
  e_i= \partial_i
  \,,
$$
thus
$\{e_\mu\}_{\mu=0}^n\subset T_pM$ is the basis dual to $\{\theta^\mu\}_{\mu=0}^n$.  Decomposing vectors $Y\in T_pM$ as $Y=Y^\alpha e_\alpha$, the equation defining $\Gammapp$ takes the form 
$$
 \forall\  \alpha_i\in \R\,,\alpha_0 \in (0,\infty)
  \quad Y^0\alpha_0 + Y^i \alpha_i \ge 0 
  \,.
$$
Clearly $Y^i=0$, $Y^0\ge0 $, and the propagation cone at $p$ is 
\begin{equation}\label{22I26.1}
\Gammapp = \{ \lambda \big(\partial_\tau -X^i\partial_i
\big)
\,,\ \lambda \in [0,\infty)
\}
\,.
\end{equation}

  \item Suppose instead  that $N$ does not vanish  at $p\in M$. Then  $\mcTpp$ is the connected component of the quadratic open cone 
\begin{eqnarray} 
  \mcTp  &  =&  
   \{ (\myxitau,\myxi)\in T^*_pM: (\myxitau  -  X^\ell\myxi_\ell)^2 > N^2 h^{ij}\myxi_i\myxi_j \} 
\end{eqnarray}
on which $\myxitau-X^\ell\myxi_\ell>0$. 
By elementary Lorentzian geometry, the propagation cone $\Gammapp$ is the open cone of timelike future-directed vectors for the metric  
\begin{equation}\label{22I25.21} 
 -d\tau^2 + N^{-2}h_{ij}(dy^i + X^i d\tau) (dy^j + X^j d\tau) 
 \,;
\end{equation} 
equivalently, for its conformal rescaling 
\begin{equation}\label{22I25.22} 
 - N^2 d\tau^2 +  h_{ij}(dy^i + X^i d\tau) (dy^j + X^j d\tau) 
 \,.
\end{equation} 
\end{enumerate}

We note that the causality properties of the equations considered here fit well into the analysis of closed cone geometries of Minguzzi~\cite{MinguzziClosedCones}. In the terminology there, the cones are proper at points at which $N\ne 0$.

To proceed, some terminology will be needed. 

\begin{Definition}
 \label{D9vi25.1}
 Let $I$ be an interval and let $\gamma$ be a locally Lipschitz curve   $\gamma:I\to \R\times \hyp$. 
 \begin{enumerate}
 \item
  We shall say that $\gamma$ is 
$\star$-causal if its field of tangents  (which is defined Lebesgue-almost everywhere by a theorem of Rademacher), lies in the propagation cone, and is non-vanishing, for Lebesgue-almost-all values of the parameter in  $I$. 
 
\item
 We say that a set $\mcU$ is $\star$-acausal if there are no $\star$-causal  curves between points of $\mcU$.

\item
The domain of dependence of a  $\star$-acausal set $\mcU$ is defined as the set of points $p \in\R\times \hyp$ such that every 
inextendible $\star$-causal curve through $p$ intersects $\mcU$ precisely once. 

\item 
We shall say that a Lorentzian metric $\hat g$  \emph{is larger}  on a set $\mcU$
if for every point $p\in\mcU$ the propagation cone, as defined above, is a subset of the set of  timelike future directed vectors for $\hat g$.

\item
We will say that 
an open subset $\mcU$ of $\R\times\hyp$ is $\star$-globally hyperbolic with respect to $\{\tau=0\}$, or simply $\star$-globally hyperbolic, if there exists a \emph{larger} differentiable Lorentzian metric $\hat g$  on $\mcU$ with the property   
that the set of $\hat g$-causal curves through $\mcU$ intersects  $\{\tau=0\}$ in a compact set, each curve precisely once. 
Equivalently, $\mcU$ is globally hyperbolic for the metric $\hat g$, with Cauchy surface $\mcU\cap \{\tau = 0 \}$.
\hfill$\Box$
\end{enumerate}
\end{Definition}
  
Since
$$
 (g_{\mu\nu} dx^\mu dx^\nu)^\star(\partial_\tau- X^i\partial_i,\partial_\tau- X^i\partial_i) = - N^2
 \,,
$$
where $(g_{\mu\nu} dx^\mu dx^\nu)^\star$ is given by \eqref{ADM},  
we see  that for every larger metric $\hat g$ the vector field  $\partial_\tau- X^i\partial_i$ is $\hat g$-timelike and future directed everywhere.

\begin{Remarks} \label{R19IV25.1}
{\rm

a)
By standard causality theory, $\star$-global hyperbolicity coincides with the usual global hyperbolicity when $N$ has no zeros. 

b)
Alternatively, one could define $\star$-global  hyperbolicity using directly causal curves as determined by our symmetric hyperbolic system. We have found the possibility to open slightly the light-cones useful for the nonlinear equations at hand; we do not know whether or not this weaker definition would suffice for our purposes below. With our stronger requirement it suffices to appeal to the standard theory of   causality associated with a Lorentzian metric, whenever needed.  
}
\hfill$\Box$
\end{Remarks}
  \subsection{Constraints}
 \label{ss8VIII25.2}

All of the work needed to analyse causality for the system of equations governing the 
propagation of the constraints has already been done in Section~\ref{ss5VIII25.1} and in the last section.
From what has been said the characteristic polynomial of the system is
\begin{equation}\label{8VIII21.11}
  p(\omega,\myxi) = (\omega-X^\ell k_\ell)^{K-2n} (\omega-X^\ell k_\ell - N|k|_h)^{n}
   (\omega-X^\ell k_\ell + N|k|_h)^{n}
    \,,
\end{equation} 
where $K$ is given by \eqref{8VIII25.12}. The causality properties of the constraints-system are identical to these of the Anderson-York system.
In particular, for smooth initial data with constraints vanishing on a set $\Omega\subset \hyp$, the vanishing of the constraints in the $\star$-domain of dependence of  $\Omega$ (and hence throughout any $\star$-globally hyperbolic development of $\hyp$)  follows from Theorem~1.2 of \cite{Rauch2}. 

Since \cite[Theorem~1.2]{Rauch2} assumes smoothness of all fields involved, an argument 
 for  data of Sobolev-class is  in order. Indeed, we claim that  an identical result on the vanishing holds by a  density-and-exhaustion argument: Consider a vacuum  set $(\hyp,N,X^i,h_{ij},K_{ij})$ of sufficiently high Sobolev-class, as needed for a well-posed propagation of the constraints, and let $(\hyp,N(k),X(k)^i,\hat h(k)_{ij},\hat K(k)_{ij})$, $k
\in
N$ be any  sequence of smooth fields approaching $(\hyp,N,X^i,h_{ij},K_{ij})$. Let $p\in \hyp$, we can use the conformal method on a sufficiently small neighborhood $\mcO\subset \hyp$ of $p$ to correct 
$(\hat  h(k)_{ij},
\hat K(k)_{ij})$ so that the fields  
$(\mcO,h(k)_{ij},K(k)_{ij})$ satisfy the vacuum constraints and converge to   $(\mcO,h_{ij},K_{ij})$. By Cauchy stability the maximal solutions of the Anderson-York equations with data $(\mcO,N(k),X(k)^i,h(k)_{ij},K(k)_{ij})$, with  $\chi(k)_{ijk}|_{\mcO}$ chosen so that $\cC(k)_{ijk}|_{\mcO} =0$, converge to the solutions of the Anderson-York equations with data $(\mcO,N ,X ^i,h _{ij},K _{ij})$, with  $\chi_{ijk}|_{\mcO}$ chosen so that $\cC_{ijk}|_{\mcO} =0$,
on any compact subset of the domain of dependence, call it $\mcD$, of these last data. As the approximating sequence has vanishing constraint fields on $\mcD$, so will have the limiting one. A simple covering argument finishes the proof.  
 
\section{Existence, uniqueness}
 \label{s22IV25.11}
 
From what has been said it follows that the equations considered  describe a field with finite speed of propagation. Therefore a solution of the Cauchy problem can be patched together from local solutions by standard arguments; in particular no restrictions on the asymptotic behaviour of the initial data are needed. 

Solutions of the Anderson-York equations, obtained from initial data on $\hyp \approx \{0\}\times \hyp$, are defined on subsets of $\R \times \hyp$.  
Standard arguments~\cite{SbierskiMGHD,ChBGeroch,EperonReallSbierski,ChMGHD}  
  adapt easily to the situation at hand, and show that for any sufficiently differentiable  density $Q$ on $\R \times \hyp$, vector field $X$ on $\R \times \hyp$, and initial data $(h_{ij},K_{ij})|_{\tau=0}$ on $\hyp$, there exists a unique maximal $\star$-globally hyperbolic   
  subset of $\R\times\hyp$ on which there exists a unique solution   $h_{ij}$ which is smooth and Riemannian. 
  Note that the non-uniqueness observed in~\cite{EperonReallSbierski} does not occur here 
   as $\partial_\tau-X^i\partial_i$ 
    is $\star$-causal everywhere (compare~\cite[Equation~(4.62)]{EperonReallSbierski}).

A very rough estimate of the norms needed is as follows: Let $k$ be the smallest integer larger than $n/2$, 
and let
$$
 \N\ni s \ge k + 2
  \,.
$$
Suppose that 
\begin{equation}\label{19IV25.1}
 Q \in \cap_{i=0}^k C^i
  \big(\R,H^{s+k + 1  -i}_\loc(\hyp)
  \big)
 \,,
 \ Y^i \in \cap_{i=0}^k C^i
  \big(\R,H^{s+ k  -i}_\loc(\hyp)
  \big) 
  \,.
\end{equation}
Assume that $h_{ij}|_{\tau=0} $ is Riemannian, with
\begin{equation}\label{19IV25.2}
 h_{ij}|_{\tau=0}  \in H^{s }_\loc(\hyp) 
 \,,
 \quad
 K_{ij}|_{\tau=0}  \in H^{s-1 }_\loc(\hyp) 
  \,.
\end{equation}
For any such data
 there exists  a unique maximal solution  of \eqref{Dvh'a}-\eqref{DvKa} 
 defined on a unique $\star$-globally hyperbolic subset of $\R\times \hyp$, with differentiability class on each $\tau$-slice as in \eqref{19IV25.2}.

One checks that the above regularity conditions more than suffice to guarantee that the vanishing of $(\mcC,\mcC_i)$ at $\tau=0$  propagates.

It is  clear that the result remains true under less stringent conditions on the data, which can be proved by a careful inspection of the terms arising in the equations, but we have not attempted to carry out the details of this.   


 The possibility of continuing  the solution is obstructed by the $H^{s}_\loc \times H^{s-1}_\loc \times L^\infty_\loc$ norm for $(h_{ij},K_{ij}, h^{ij})$, and so finiteness of this norm provides a local continuation criterion  for the solutions. 
 
\section{Examples}
  
It is instructive to have in mind a few examples when considering our slicings.

The simplest example is provided by $N\equiv 0$, in which case the ``slicing'' consists of a single slice. The evolution equations become
\begin{equation}\label{10IV25.2}
(\partial_\tau - \mathcal{L}_X) h_{ij} = 0= 
(\partial_\tau - \mathcal{L}_X) K_{ij}  
 \,.
\end{equation}
Thus both the metric and the extrinsic curvature tensor evolve according to the flow of $X$. Note that, on non-compact manifolds,
the flow of most vector fields   runs away to infinity in finite time, in which case the solution will only be defined on a proper subset of $\R\times \hyp$.

As another simple example, let $(\mcM,g)$ be any smooth  
spacetime  
equipped with a  
time function $t$, so that we can write $\mcM = \R \times \hyp$, where the level sets of $t$ are defined by 
the projection on the $\R$ factor. Letting the coordinates $y$ to be Lie-dragged along the normals to the level sets of $t$, one can write the metric $g$ in the ADM form with no shift:
\begin{equation}\label{6IV25.1a}
 g= 
 - \mathring N^2 dt^2 + \mathring h_{ij}  dy^i  dy^j  
  \,,
\end{equation}
where both $\mathring N$ and $\mathring h$ depend upon both $t$ and $y^i$ in general.
Set
 $t= \tau^2$, so that the metric $g$ equals 
\begin{equation}\label{6IV25.1b}
 g= 
 - 4\mathring N^2 \tau^2  d\tau^2 + \mathring h_{ij}  dy^i  dy^j  
  \,,
\end{equation}
which is of the ADM form \eqref{ADM0} with 
\begin{equation}\label{6IV25.1c}
  N (\tau, y) = 2 \tau \mathring N(t=\tau^2,y)\,,
  \qquad
    h_{ij} (\tau,y) =  \mathring h_{ij}(t=\tau^2,y)
    \,.
\end{equation}
Note that going forward in $\tau$ corresponds to going backwards in $t$ for $\tau\le 0$, and forward in $t$ otherwise. We emphasise  that the spacetime metric at a given $t$ does not depend upon whether $\tau$ is positive or negative. 
The set $\{t>0\}$ is covered twice by the  evolution in $\tau$-time, and
 returning to the same spacetime point with distinct values of $\tau$ one experiences an identical spacetime metric. This fact follows on general grounds, without knowing the explicit form of the metric  above, from the fact that the Cauchy problem for the Einstein equations on a smooth spacelike hypersurface has unique solutions up to isometry, and that the Anderson-York equations have unique solutions; more on this in Section~\ref{s13IV25.1} below.

Our next example is that of two
``Rindler wedges'' attached together:
\begin{equation}\label{8IV23.1}
   -(y^1)^2   d\tau^2 +     \delta_{ij}dy^i dy^j 
   \,.
\end{equation}
This tensor field is obtained by moving the hypersurface $\{t=0\}$ in Minkowski space-time by a boost along the first axis. We have $N=y^1$, so that the evolution is forward in Minkowskian time for $y^1>0$, is frozen at $y^1=0$,  and proceeds backwards in Minkowskian time for $y^1<0$. 

An example with a similar flavour is provided by the spacetime version of the Einstein-Rosen bridge. Indeed, replacing $r$ in the standard form of the exterior Schwarzschild metric by a new  coordinate $\rho$  given by
$$ \rho= \sqrt{r^2-4m^2}\ \Longleftrightarrow \ r =
\sqrt{\rho^2+4m^2}
\,,
$$ 
the Schwarzschild metric becomes 
\begin{equation}\label{mygoodcord} 
-\Big(1 - \frac{2m}{\sqrt{\rho^2+4m^2}}\Big) dt^2  +  \Big(1+
\frac{2m}{\sqrt{\rho^2+4m^2}}\Big) d\rho^2 + (\rho^2+4m^2)
d\Omega^2\,.\end{equation}
The $t$-independent lapse function 
$$
N = \sqrt{1 - \frac{2m}{\sqrt{\rho^2+4m^2}}} \ge 0 
$$
smoothly extends through its zero-set   $\{\rho=0\}$. Here the evolution proceeds forward in $t $,  with a lapse function which is positive away from $\{\rho=0\}$. 
This appears surprising until one realises that the constant-$t$ slicing of the Kruskal-Szekeres extension of the Schwarzschild spacetime is determined by the flow of the static Killing vector, and this Killing vector changes time-orientation when crossing the Killing horizon across its bifurcation surface. So the slicing progresses forward in $t$, which corresponds to opposite time-orientations in the Kruskal-Szekeres manifold when crossing $\rho=0$.

\section{Connecting with the maximal globally hyperbolic development}
\label{s8IV25.1}
 
Let $(\mcM, g_{\mu\nu})$ be the maximal globally hyperbolic vacuum development (MGHD) of a vacuum initial data set   $(\hyp,h_{ij}(0,y),K_{ij}(0,y))$. Let 
$$
 (h_{ij}(\tau,y),K_{ij}(\tau,y),N(\tau,y),X^i(\tau,y))
$$
satisfy the {Anderson-York} evolution equations on an open subset of $\R\times \hyp$ containing the initial data surface $\{0\}\times \hyp$. The question arises, how does an  ADM metric obtained by evolving the {Anderson-York} equations embed into   $(\mcM, g_{\mu\nu})$. The aim of what follows is to explore this question.

 By definition of a maximal globally hyperbolic development (MGHD), say $(\mcM,g)$, of a data set $(\hyp,h_0,K_0)$,  
 for every globally hyperbolic development $(\mcM_0,g_0)$ thereof  there exists an isometric embedding $\phi:\mcM_0\to\mcM $  which preserves the initial data, in the sense that
 there exists an embedding $\phi^\mu(0,y)$ of $\hyp$ into $\mcM$ so that
 $h_{ij}(0,y)$ is the pull-back to $\hyp$ of the induced metric on $\hyp$ and $K_{ij}(0,y)$ is the second fundamental form
 (see \cite{SbierskiMGHD,ChMGHD,RingstroemBook,SCC,EperonReallSbierski} for more on this).  

We note the following elementary observation:

\begin{Proposition}
  \label{P2VI26.1}
Let $(Q,X^i)$ be given on $\R\times \hyp$ and let $\hyp_0$ denote  
the open subset of $\hyp$ on which $Q$ has no zeros. 
For any $\star$-globally hyperbolic vacuum development
$$
 \mcM_0\subset \R\times \hyp
$$
 of  the data  $(\hyp_0,h_{ij}(0,y),K_{ij}(0,y))$ on which $Q$ has no zeros
there exists an embedding $\phi$ of $\mcM_0$ into the maximal globally hyperbolic development $(\mcM,g)$ of the data such that
\begin{equation}\label{ADM2b}
\phi^*g = -  N^2 d\tau^2 + h_{ij} (dy^i +  X^i d \tau) (dy^j +  X^j d \tau)\,.
\end{equation}
\end{Proposition}

\begin{remark}
  \label{R2VI25.1}
{\rm  
  The restriction on the zeros of $Q$ will be removed in the next section. 
  }
  \hfill$\Box$
\end{remark}

{\noindent\sc Proof:} The ADM metric \eqref{22I25.22} restricted to $\mcM_0$ is Lorentzian and vacuum, and the $\star$-globally hyperbolic development is globally hyperbolic in the usual sense for the  metric  \eqref{22I25.22}. 
It follows from the defining property of a MGHD that there exists an isometric embedding 
$$
 \phi:\mcM_0\to\mcM
 \,.
$$
In particular the image of each slice
$$
 \hyp_\tau:= \phi\big((\{\tau\}\times \hyp)\cap \mcM_0\big)
$$
is an embedded, not necessarily connected, hypersurface in $\mcM$. 
Let us define a lapse function $\bar N$ and a shift vector $\bar Y$ by the formula \eqref{lapse-shift}
\begin{equation}\label{lapse-shifta}
\dot{\phi}^\mu (\tau,y) = \bar N (\tau,y) 
\, \bar n^\mu\big(\phi (\tau,y)\big) + \phi^\mu{}_{{, i}} (\tau,y) \bar X^i (\tau,y)
\,,
\end{equation}
where  $\bar n^\mu  $ denotes the field of future-directed unit normals to the smooth spacelike hypersurfaces $\phi^\mu(\{\tau\}\times \hyp)$. 
Here for each $\tau$ the field $\bar n^\mu$ is a vector field defined along  $\hyp_\tau$,  
and defines in an obvious way a vector field on the image $\phi(\mcM_0) \subset \mcM$ of $\mcM_0$ by $\phi$.
The calculation leading to \eqref{ADM} gives
\begin{equation}\label{ADM2c}
(g_{\mu\nu} dx^\mu dx^\nu)^\star = - \bar N^2 d\tau^2 + \bar h_{ij} (dy^i + \bar X^i d \tau) (dy^j + \bar  X^j d \tau)\,,
\end{equation}
for some Riemannian metric $\bar h_{ij}$. 
Since $\phi$ is an isometry we conclude that 
\begin{equation*} 
  \bar h_{ij} = h_{ij}
  \,,
  \quad
  \bar N = N
  \,,
  \quad
  \bar X^i =X^i
  \,.\tag*{$\Box$}
\end{equation*}
\medskip

 What happens away from $\mcM_0$ is   clear in some simple cases.
Suppose, for instance, that the set of zeros of the lapse function is $\tau$-independent. Let us denote by $\Omega\subset \hyp$  the set of points where $N$ vanishes. On $\R \times \Omega$ there is no motion of the slices $\phi(\hyp)$ in the associated maximal globally hyperbolic spacetime: indeed, for $y\in \Omega$ we can set
\begin{equation}\label{8IV25.1}
 \phi (\tau,y)= \phi(0,y)
 \,.
\end{equation}
On $\R\times \Omega$ the evolution equations \eqref{Dvh'a}-\eqref{DvKa} become
\begin{equation}\label{8IV25.2}
(\partial_\tau - \mathcal{L}_X) h_{ij} = 0
\,,
\end{equation}
\begin{equation}\label{8IV25.3}
(\partial_\tau - \mathcal{L}_X) K_{ij} =   D_i D_j N
 \,,
\end{equation}
so that the metric moves by the flow of (a possibly $\tau$-dependent) shift vector $X$. (As already pointed-out,  the flow will escape to infinity in finite time for most vector fields $X$, so strictly speaking the above only applies to the subset of $\R\times\Omega$ on which this flow is defined.)  On the interior of $\Omega$, if any, the Hessian of $N$ vanishes, and $K_{ij}$ also evolves there according to the flow of $X$. 

Away from the set $\{(\tau,y^i)\in \R\times\Omega\}$ we are in the globally hyperbolic setting just discussed, and the embedding 
$$
 \phi:
  \underbrace{
   \mcM_0}_{
   \subset \R\times (\hyp\setminus\overline\Omega)}
   \to \mcM
$$ 
extends by continuity to  \eqref{8IV25.1} on $\R\times \partial \Omega$. This provides a complete description of the globally hyperbolic part of the solution of the Anderson-York equations, as seen by the maximal globally hyperbolic development of the data,  in the case of a lapse with a $\tau$-independent zero set.

Next, consider the situation where all the fields $(h_{ij}|_{\tau =0}, K_{ij}|_{\tau =0}, N, X^i)$ 
are real-analytic. Then the solution of the Anderson-York equations will be real-analytic by propagation of analyticity by symmetric-hyperbolic systems. The maximal globally hyperbolic development $(\mcM,\fourg)$ will likewise be real-analytic, and the embedding equation 
\begin{equation}
(\partial_\tau - X^l \partial_l) \phi^\mu = N n^\mu
\,,
\end{equation}
with $n^\mu$  given by \eqref{4V25.1} when $n=3$, or by the obvious generalisation thereof in other dimensions, can  be solved by the Cauchy-Kovalevskaya theorem to provide a slicing of $(\mcM,\fourg)$.

\section{A symmetrisable-hyperbolic system for slicings}
 \label{s13IV25.1}

We are ready now to pass to the proof of the following; for simplicity we assume smoothness of the fields involved, but the argument holds with data of sufficiently high local Sobolev regularity:

\begin{theorem}
 \label{T2VI25.1}
Consider a smooth solution of the Anderson-York equations defined on a $\star$-globally hyperbolic subset $\mcU$ of $\R\times \hyp$ 
evolving out of  vacuum initial data on $\hyp$.
There is a map  $\phi $ of $\mcU$  into the maximal globally hyperbolic vacuum development of the initial data on $\{0\}\times \hyp$ such that each connected component of the image  $\phi(\tau, \cdot)$ is a smooth embedded spacelike submanifold of $\mcM$.  
\end{theorem}
 
\begin{remark}
{\rm
We note that the images by  $\phi(\tau, \cdot)$ of $\mcU$ need not be connected because the intersections of $\mcU$ with the surfaces of constant $\tau$ will not be connected in general.
}
\hfill$\Box$
\end{remark}
 
\begin{remark}
{\rm
In particular any sets 
$$
\mbox{$\big(Q,X^i, h_{ij}(0,\cdot), K_{ij}(0,\cdot)\big)$ and  $\big(
\hat Q,
\hat X^i, h_{ij}(0,\cdot), K_{ij}(0,\cdot)\big)$ }
$$
with $(Q,X^i)\ne (\hat Q,\hat X^i)$   can both be obtained by a slice in  the same spacetime. 
\hfill$\Box$
}
\end{remark}
 
We start with the 

\medskip

{\noindent\sc Proof of Theorem~\ref{T29V25.1}:} 
Let us assume that the initial slice $\phi(0,\cdot)\hyp \subset \mcM$ is contained in a Cauchy surface in a globally hyperbolic spacetime $(\mcM,g)$.
 In the context of Theorem~\ref{T2VI25.1} this
 is obtained by taking $(\mcM,g)$ to be the maximal globally hyperbolic development of the data, and 
involves no loss of generality 
for Theorem~\ref{T29V25.1}, where the claim is local.

The idea is to construct a 
symmetrisable hyperbolic system  
associated with  the  equation for slicings, namely
\begin{equation}
(\partial_\tau - X^\ell\partial_\ell ) \phi^\mu = N n^\mu\,;
\label{13IV25.61}
\end{equation} 
recall that $n^\mu$ is an explicit function of the spacetime metric, of $\phi^\mu$  and of $\phi^\mu{}_{,i}$.
Instead of using the unit-normal $n^\mu$ it turns out to be useful to consider the rescaled normal $m^\mu$, defined as
\begin{equation}
m^\mu = (\mathrm{det}\, h_{ij})^\frac{1}{2} n^\mu\,,
\end{equation}
where 
\begin{equation}\label{19IV25.5}
 h_{ij} = \phi^\mu{}_{,i}\phi^\nu{}_{,j}g_{\mu\nu}(\phi )
 \,.
\end{equation}
It holds that
\begin{equation}\label{star}
D_v m^\mu = (\mathrm{det}\, h_{ij})^\frac{1}{2}D^i (N \phi^\mu{}_{,i})\,.
\end{equation}
For the proof of this last equation,  recall that
\begin{equation}
D_v n^\mu = \phi^\mu{}_{,i} D^i N\,.
\end{equation}
Thus 
\begin{equation}
D_v ((\mathrm{det}\, h_{ij})^\frac{1}{2} n^\mu) 
= (\mathrm{det}\, h_{ij})^\frac{1}{2}(\phi^\mu{}_{,i} D^i N + N K n^\mu)\,.
\end{equation}
Finally observe, from $D_i \phi^\mu{}_{,j} = n^\mu K_{ij}$, that
\begin{equation}
D^i \phi^\mu{}_{,i} = K n^\mu = K (\mathrm{det}\, h_{ij})^{-\frac{1}{2}} m^\mu\,.
\end{equation}
The result follows.  

We now have, in addition to 
\begin{equation}
(\partial_\tau - X^\ell\partial_\ell ) \phi^\mu = (\mathrm{det}\, h_{ij})^{-\frac{1}{2}} N m^\mu\,,
\end{equation} 
the equation (\ref{star}) and 
\begin{equation}
D_v \phi^\mu{}_{,i} = D_i ((\mathrm{det}\, h_{ij})^{-\frac{1}{2}} N m^\mu)\,,
\end{equation}
which is obtained by $i$-differentiating the last equation.
After introducing $f^\mu{}_i = \phi^\mu{}_{,i}$, we obtain the system 
\begin{eqnarray}
   \label{12IV25.7}
D_v\phi^\mu 
 & = & 
  Q   m^\mu
 \,,
  \\
D_vf^\mu{}_i  
 & = & D_i (Q  m^\mu)  
\,,
\\
D_v m^\mu 
 &= & (\mathrm{det}\, h_{ij})^\frac{1}{2}D^i (  (\mathrm{det}\, h_{ij})^{\frac{1}{2}} Q f^\mu{}_{i}) 
 \,.
   \label{12IV25.5}
\end{eqnarray} 
 Here $D_i$ is the horizontal covariant derivative with the additional convention that terms containing $\Gamma^{\mu}_{\nu\rho}(\phi)\phi^\rho{}_{,i}$ are replaced by $\Gamma^{\mu}_{\nu\rho}(\phi)f^\rho{}_i$. 
 This system is clearly symmetrisable if one ignores the fact that 
 derivatives of $h_{ij}$,  
which involve derivatives of $f^\mu{}_i$ in view of \eqref{19IV25.5}, occur at the right-hand side. 
This can be cured by adding to the fields $(\phi^\mu,m^\mu,f^\mu{}_i)$ the fields  $ (h_{ij}, K_{ij}, \chi_{ijk} ) $ of Section~\ref{s2III25.1}, as follows.

For each occurrence of 
$$ h_{ij} \equiv \phi^\mu{}_{,i}\phi^\nu{}_{,j}g_{\mu\nu} \big(\phi(\tau,y)\big)
$$ 
in the right-hand side of the equations \eqref{12IV25.7}-\eqref{12IV25.5} we write $\bar h_{ij}$. Similarly, 
for each occurrence of $\zD_{k} h_{ij}$ in these equations  we write
\begin{equation}
  \bar \chi_{ijk}  + 4 \bar h_{k(i} \bar \chi_{j)}
  \,, 
 \quad
  \mbox{ where } 
  \bar \chi_i  := \frac{1}{n-2}\bar h^{jk}\bar \chi_{j[ki]} 
  \,.
   \label{22IV25.9a}
\end{equation}
We complement the   equations so obtained with the following version of the evolution equations \eqref{4II25.21}-\eqref{4II25.21c}: 
First, instead of \eqref{4II25.21} we write
\begin{align}\label{4II25.21ne}
  &
(\partial_\tau - \mcL _X) \bar h_{ij} =  2 N \bar K_{ij} 
\,. 
& 
\end{align}
Next, recall \eqref{4II25.4- }, which was at the origin of \eqref{4II25.21b}, and  which we reproduce here for the convenience of the reader: 
\begin{equation}
(\partial_\tau - \mathcal{L}_X)   
 K_{ij}=   \frac{N}{2}   h^{kl}  \zD_l 
 (\chi_{ijk} - \cC_{ijk})  - N R_{ij} 
-\breve N_{ij}    + 2 N K_{i\ell} K^\ell{}_j  - NK K_{ij}  
 \,.
  \label{4II25.4-dfg}
\end{equation}
Instead of \eqref{4II25.4-dfg} we will use the equation
\begin{equation}
(\partial_\tau - \mathcal{L}_X)  \bar  K_{ij}=   
 \frac{N}{2}  \bar  h^{kl} \zD_l 
  \bar \chi_{ijk}  - N \bar R_{ij} 
-\breve{\bar N}_{ij}    + 2 N \bar K_{li} \bar K_j{}^l - N\bar K \bar K_{ij}  
 \,,
  \label{4II25.4-dfd}
\end{equation}
where in \eqref{4II25.4-dfg} we replaced $ \cC_{ijk} $ by $0$,   
and we  replaced  
\begin{eqnarray}  
 R_{ij}  (\tau,y)
  & \equiv &  
   \phi^\mu{}_{,i}\phi^\nu{}_{,j}R_{\mu\nu} \big(\phi(\tau,y)\big) 
   \label{22IV25.1bb}  
\end{eqnarray}
 by 
\begin{eqnarray}  
 \bar R_{ij}  (\tau,y)
  & := &  
    f^\mu{}_{i}f^\nu{}_{j} R_{\mu\nu}\big(\phi(\tau,y)\big)
   \label{22IV25.1b} 
  \,.
\end{eqnarray}
As before all, whether  explicit or implicit in 
\eqref{4II25.4-dfg},  
occurrences of the fields $ h_{ij}$ and $\zD_{k} h_{ij} $  are replaced by $\bar h_{ij}$ 
together with
\begin{equation}
 \zD_{k} h_{ij} \to 
  \bar \chi_{ijk}  + 4 \bar h_{k(i} \bar \chi_{j)}
  \,, 
 \quad
  \mbox{ where } 
  \bar \chi_i  := \frac{1}{n-2}\bar h^{jk}\bar \chi_{j[ki]} 
  \,.
   \label{22IV25.9}
\end{equation}
In particular the Christoffel symbols in $D$ have been replaced by the obvious expressions involving $\bar h_{ij}$ and $\bar \chi_{ijk}$ in the operator $\bar D$.

Finally, recall \eqref{Dvchiolala}, which was at the origin of \eqref{4II25.21c}:
\begin{equation}\label{22IV25.7}
(\partial_\tau - \mcL _X) \chi_{ijk} =  2 N (\zD_k  K_{ij} - 2\,h_{k(i}\mathcal{C}_{j)}) + \breve s_{ijk} +(\partial_\tau - \mcL _X) \cC_{ijk} 
 \,.
\end{equation} 
Instead we use
\begin{equation}\label{22IV25.8}
(\partial_\tau - \mcL _X) \bar \chi_{ijk} =  2 N (\bar D_k \bar K_{ij} - 2\bar h_{k(i}\bar{\cC}_{j)}) + \bar s_{ijk}  
 \,,
\end{equation} 
with substitutions as in the previous equations,  
together with the replacement of
\begin{eqnarray}  
 \cC_{i}  (\tau,y)
  &= & \phi^\mu{}_{,i}\ n^\nu  (G_{\mu\nu}+ \Lambda g_{\mu\nu})
   \big(\phi(\tau,y)\big)
\end{eqnarray}
by
\begin{eqnarray} 
 \bar \cC_{i}  (\tau,y)
   & = &
     f^\mu{}_{i}(\det \bar h_{k\ell} )^{1/2} \bar  m^\nu  (G_{\mu\nu}+ \Lambda g_{\mu\nu})\big(\phi(\tau,y)\big)
  \,.
   \label{24IV25.1}
\end{eqnarray}

We now have  a system of equations consisting of the barred-version of \eqref{12IV25.7}-\eqref{12IV25.5},
i.e.
\begin{eqnarray}
   \label{12IV25.7cde}
\bar D_v\phi^\mu 
 & = & 
  Q   \bar m^\mu
 \,,
  \\
\bar D_vf^\mu{}_i  
 & = & \bar D_i (Q  \bar m^\mu)  
\,,
\\
\bar D_v\bar  m^\mu 
 &= & (\mathrm{det}\, \bar h_{ij})^\frac{1}{2}D^i (  (\mathrm{det}\, \bar  h_{ij})^{\frac{1}{2}} Q f^\mu{}_{i}) 
 \,,
   \label{12IV25.5cde}
\end{eqnarray} 
together with \eqref{4II25.21ne}, \eqref{4II25.4-dfd}, and \eqref{22IV25.8}:
\begin{eqnarray}
   \label{22IV25.8a} 
(\partial_\tau - \mcL _X) \bar h_{ij}
 &  = &   2 N \bar K_{ij}  
 \,,
  \\
(\partial_\tau - \mathcal{L}_X)  \bar  K_{ij}
& = &    
 \frac{N}{2}  \bar  h^{kl} \zD_l 
  \bar \chi_{ijk}  - N \bar R_{ij} 
-\breve{\bar N}_{ij}   
   + 2 N \bar K_{li} \bar K_j{}^l
\nonumber
\\
 && 
 - N\bar K \bar K_{ij}  
 \label{22IV25.8c1}
 \,, 
\\
(\partial_\tau - \mcL _X) \bar \chi_{ijk} 
&= &
 2 N (\bar D_k \bar K_{ij} - 2\bar h_{k(i}\bar{\cC}_{j)}) + \bar s_{ijk}  
 \,.
   \label{22IV25.8c}
\end{eqnarray} 

As initial data for the system \eqref{12IV25.7cde}-\eqref{22IV25.8c} we take $\phi |_{\tau =0}$ to be the embedding of initial interest, $\bar h_{ij}|_{\tau =0}$ 
to be the pull-back to $\hyp$ of the metric induced on $\phi |_{\tau =0}(\hyp)$ by $g_{\mu\nu}$, and we set 
\begin{equation}
\bar m^\mu|_{\tau =0} =
 \big( (\mathrm{det}\, \bar h_{ij})^\frac{1}{2} n^\mu
 \big)|_{\tau =0}
 \,,
\end{equation}
where 
$n^\mu|_{\tau =0}$ is  the normal to the image  $\phi |_{\tau =0}(\hyp)$, we let $\bar K_{ij}|_{\tau =0}$  
 be the pull-back to $\hyp$ of extrinsic curvature of the image of $\phi |_{\tau =0}(\hyp)$, with $\bar \chi_{ijk} |_{\tau =0}$ determined from $\zD_k h_{ij}  |_{\tau =0}$ using the relation inverse to \eqref{22IV25.9}: 
\begin{equation}
\bar \chi_{ijk}|_{\tau =0} = 
 \big(\zD_{k} h_{ij} + 4 h_{k(i} \gamma_{j)}
 \big)|_{\tau =0}\,.
  \label{4II25.11c}
\end{equation}

The system \eqref{12IV25.7cde}-\eqref{22IV25.8c}  is  symmetrisable-hyperbolic,  and one checks that the causality properties of the system  remain as  in Section~\ref{s21I25.1}. 
As in Section~\ref{s22IV25.11}, for sufficiently differentiable initial data of Sobolev class on $\hyp$ there exists a maximal $\star$-globally hyperbolic subset  
$$
 \mcV\subset \R\times \hyp
$$
on which a unique solution, with a positive definite tensor field $\bar h_{ij}$,  of  \eqref{12IV25.7cde}-\eqref{22IV25.8c}  exists.
  
It remains to show that the solution so obtained provides a solution of the embedding equation \eqref{13IV25.61}. %
For this let us denote by $\X$ the collection of fields 
\begin{equation}\label{4V25.3a}
  (\phi^\mu, f^\mu{}_i, \bar m^\mu, \bar h_{ij}, \bar K_{ij}, \bar \chi_{ijk})
  \,.
\end{equation} 
For $n\to\infty$ let $(N_n,X^i_n)$ be any sequence of real-analytic fields converging to $(N,X^i)$, and let $\X_n|_{\tau=0}$ be any sequence of real-analytic initial data converging to $\X|_{\tau=0}$, both convergences being in   sufficiently high-index Sobolev-topologies, as needed for well-posedness of the relevant system of equations. Let $\mcV_n$ be the domain of definition of the maximal $\star$-globally hyperbolic solution $\X_n$ with these  data.
By propagation of analyticity for symmetric-hyperbolic systems the fields $\X_n$ are real-analytic on $\mcV_n$. 

Now, one can solve directly the embedding equation
\begin{equation}
(\partial_\tau - X_n^\ell\partial_\ell ) \phi_n^\mu = N_n n_n^\mu
\label{13IV25.61x}
\end{equation} 
locally by invoking the Cauchy-Kovalevskaya theorem, using $\phi^\mu_n|_{\tau=0}$ as initial data.  
The solutions so obtained satisfy a system identical to \eqref{12IV25.7cde}-\eqref{22IV25.8c} after the renaming
\begin{equation}\label{4V25.3-}
  (\phi^\mu, f^\mu{}_i, \bar m^\mu
   , \bar h_{ij} 
   , \bar K_{ij}
   , 
   \bar \chi_{ijk})
  \to
  (\phi^\mu, \phi^\mu{}_{,i},  (\mathrm{det}\, h_{ij})^\frac{1}{2} n^\mu,   h_{ij} 
  ,     K_{ij}
  ,   
 \chi_{ijk}
 )
  \,,
\end{equation} 
with $\chi_{ijk}$ replacing $\zD_{k} h_{ij}$ using \eqref{9III25.12cde}.
Uniqueness of solutions of the  system \eqref{12IV25.7cde}-\eqref{22IV25.8c} shows that $\phi_n^\mu$ solves  the embedding equation \eqref{13IV25.61x} on $\mcV_n$.
Passing to the limit $n\to\infty$ and invoking Cauchy-stability we conclude that $\phi^\mu$ solves the embedding equation \eqref{13IV25.61} on $\mcV$, with 
\begin{equation}\label{4V25.3b3}
  (\phi^\mu, f^\mu{}_i, \bar m^\mu
   , \bar h_{ij} ,
   \bar K_{ij}, 
   \bar \chi_{ijk})
  =
  (\phi^\mu, \phi^\mu{}_{,i},  (\mathrm{det}\, h_{ij})^\frac{1}{2} n^\mu,   h_{ij},
 K_{ij},  
   \chi_{ijk})
 \,.
\end{equation} 
This establishes Theorem~\ref{T29V25.1}.
\hfill$\Box$

\medskip

{\noindent\sc Proof of Theorem~\ref{T2VI25.1}:}
In vacuum, it now follows from \eqref{4V25.3b3} that $\bar R_{ij} = \lambda \bar h_{ij}$, while \eqref{24IV25.1} shows that $\bar \cC_i=0$. 
So  
the last three equations \eqref{22IV25.8a}-\eqref{22IV25.8c}  decouple from the first three equations in  
\eqref{12IV25.7cde}-\eqref{22IV25.8c}, and can be solved independently from the first three. The equations \eqref{22IV25.8a}-\eqref{22IV25.8c} satisfied by the  fields $(\bar h_{ij},\bar K_{ij},\bar \chi_{ijk})(\tau,\cdot)$, are identical to the Anderson-York equations \eqref{4II25.21}-\eqref{4II25.21c} satisfied by  $(  h_{ij}, K_{ij},\chi_{ijk})(\tau,\cdot)$, with the same initial data. Since solutions to these equations are unique in $\star$-domains of dependence,  it holds on $\mcV$ that
$$(\bar h_{ij},\bar K_{ij})(\tau, y) \equiv  (  h_{ij},  K_{ij}) (\tau, y)
\,,
$$
where  the right-hand side denotes the fields obtained by solving the Anderson-York equations.
 This ends the proof of Theorem~\ref{T2VI25.1}.
 \hfill
 $\Box$
%

\medskip

Again in vacuum, it should be clear that the maximal $\star$-globally hyperbolic subset  $\mcV\subset \R\times \hyp$ of existence of solutions of 
\eqref{12IV25.7cde}-\eqref{22IV25.8c}, with a Riemannian $\bar h_{ij}$, coincides with the maximal $\star$-globally hyperbolic subset  $\mcU\subset \R\times \hyp$ of existence of solutions of \eqref{4II25.21}-\eqref{4II25.21c} with a Riemannian $h_{ij}$. 

We emphasise once again that the above holds regardless of existence, or not, of zeros of $N$.   

Analogous results hold for models with matter fields satisfying well behaved evolution equations.
 
It is curious, and quite unsatisfactory, that we had to introduce so many auxiliary fields to establish existence of solutions of the embedding equation~\eqref{13IV25.61}. It is tempting to conjecture that there exists a simpler set of equations which allows us to solve~\eqref{13IV25.61} without assuming analyticity of the fields involved.

\bigskip
 
{\noindent\sc Acknowledgements:}
We acknowledge questions from Thomas Baumgarte, which encouraged us to carry-out this work. 
We are grateful to   Hamed Barzegar for useful comments on a previous version of this manuscript, and to David Hilditch for useful correspondence.

%
%
%

%
%
%
%
%


\begin{thebibliography}{10}

\bibitem{anderson1999fixing}
Arlen Anderson and James~W York~Jr.
\newblock Fixing {{Einstein}'s} equations.
\newblock {\em Phys.\ Rev.\ Lett.}, 82(22):4384, 1999.

\bibitem{kidder2001extending}
Lawrence~E Kidder, Mark~A Scheel, and Saul~A Teukolsky.
\newblock Extending the lifetime of 3d black hole computations with a new
  hyperbolic system of evolution equations.
\newblock {\em Physical Review D}, 64(6):064017, 2001.



\bibitem{FournodavlosLuk}
Grigorios~Fournodavlos and Jonathan~Luk, \newblock {{Asymptotically Kasner-like singularities}},
  {\em Amer.\ Jour.\ Math.}, {145}:1183--1272, (2023).
  arXiv:2003.13591
  [gr-qc]. 

\bibitem{fournodavlos2021initial}
Grigorios Fournodavlos and Jacques Smulevici.
\newblock The initial boundary value problem for the {Einstein} equations with
  totally geodesic timelike boundary.
\newblock {\em Communications in Mathematical Physics}, 385(3):1615--1653,
  2021.
  
  
\bibitem{Hilditch:2024nhf}
David Hilditch, \emph{{Solving the Einstein equations numerically}}, {New
  frontiers in GRMHD simulations} (Cosimo Bambi, Yosuke Mizuno, Swarnim
  Shashank, and Feng Yuan, eds.), Springer Nature, 2025.
  

\bibitem{CJK}
Piotr~T Chru\'{s}ciel, Jacek Jezierski, and Jerzy Kijowski.
\newblock {\em {H}amiltonian field theory in the radiating regime}, volume m70
  of {\em Lect. Notes in Physics}.
\newblock Springer, Berlin, Heidelberg, New York, 2001.
\newblock URL
  \url{http://www.phys.univ-tours.fr/~piotr/papers/hamiltonian_structure}.
  
  

\bibitem{PSchmidt}
Patricia Schmidt.
\newblock On a general 3+1 formalism: electrodynamics and ideal hydrodynamics
  in curved spacetime.
\newblock Diploma thesis, University of Vienna, 2010.


\bibitem{baumgarte2022shock}
Thomas~W Baumgarte and David Hilditch.
\newblock Shock-avoiding slicing conditions: Tests and calibrations.
\newblock {\em Phys.\ Rev.\ D}, 106:044014, 2022.



\bibitem{marsden1994mathematical}
Jerrold~E Marsden and Thomas~JR Hughes.
\newblock {\em Mathematical foundations of elasticity}.
\newblock Courier Corporation, 1994.



\bibitem{ericksen1960appendices}
Clifford Truesdell and Richard A. Toupin. 
\newblock 
The Classical Field Theories. 
\newblock
In: Flügge, S. (eds) Principles of Classical Mechanics and Field Theory / Prinzipien der Klassischen Mechanik und Feldtheorie. Encyclopedia of Physics / Handbuch der Physik, vol 2 / 3 / 1. Springer, Berlin, Heidelberg. Jerald Ericksen, 1960.
  
\bibitem{Clebsch}
Alfred Clebsch.
\newblock  Ueber eine Fundamentalaufgabe der Invariantentheorie. 
\newblock {\em Math.\ Ann.}, 5:427–-434, 1872.  

\bibitem{eells1964harmonic}
James Eells and Joseph~H Sampson.
\newblock Harmonic mappings of {Riemannian} manifolds.
\newblock {\em Am.\ Jour.\ of Math.}, 86(1):109--160, 1964.


\bibitem{gourgoulhon20123+}
Eric Gourgoulhon.
\newblock {\em 3+ 1 formalism in general relativity}.
\newblock Springer, 2012.


\bibitem{giulini2014dynamical}
Domenico Giulini.
\newblock Dynamical and {Hamiltonian} formulation of general relativity.
\newblock {\em Springer Handbook of Spacetime}, pages 323--362, 2014.


\bibitem{Frittelli}
Simonetta Frittelli.
\newblock {Note on the propagation of the constraints in
  standard 3+1 general relativity}, 
\newblock {\em  Phys.\ Rev.\ D} {55:5992--5996}, 1997.



\bibitem{Taylor91}
Michael E. Taylor 
\newblock {Pseudodifferential Operators and nonlinear PDE}.
\newblock {Birkh{\"a}user Boston},
 {1991}.


\bibitem{JMR}
Jean-Luc Joly, Guy M\'etivier, and Jeffrey Rauch.
\newblock Hyperbolic domains of determinacy and {H}amilton-{J}acobi equations.
\newblock {\em J. Hyperbolic Differ. Equ.}, 2:713--744, 2005.

\bibitem{Rauch1}
Jeffrey Rauch.
\newblock Precise finite speed with bare hands.
\newblock {\em Methods Appl. Anal.}, 12(3):267--277, 2005.

\bibitem{Rauch2}
Jeffrey Rauch.
\newblock Precise finite speed and uniqueness in the {C}auchy problem for
  symmetrisable hyperbolic systems.
\newblock {\em Trans. Amer. Math. Soc.}, 363(3):1161--1182, 2011.

 
\bibitem{MetivierSym}
Guy.~M\'{e}tivier.
\newblock {{$L^2$} well-posed {C}auchy problems and
  symmetrisability of first order systems}.
\newblock  {\em Jour.\ \'{E}c. Polytech. Math.},
   {1}:39--70, 2014.  

\bibitem{ChBlackHoles}
Piotr~T Chru\'{s}ciel.
\newblock {\em Geometry of Black Holes}.
\newblock Oxford University Press, 2020.

\bibitem{MinguzziClosedCones}
Ettore Minguzzi.
\newblock {Causality theory for closed cone structures with applications}.
\newblock {\em Rev.\ Math.\ Phys.}, 31:1930001, 139, 2019.
\newblock arXiv:1709.06494 [gr-qc].

\bibitem{SbierskiMGHD}
Jan Sbierski.
\newblock On the existence of a maximal {C}auchy development for the {E}instein
  equations: a dezornification.
\newblock {\em Ann.\ Henri Poincar\'e}, 17:301--329, 2016.
\newblock arXiv:1309.7591 [gr-qc].

\bibitem{ChBGeroch}
Yvonne Choquet-Bruhat and Robert Geroch.
\newblock Global aspects of the {C}auchy problem in general relativity.
\newblock {\em Commun.\ Math.\ Phys.}, 14:329--335, 1969.

\bibitem{EperonReallSbierski}
Felicity~C. Eperon, Harvey~S. Reall, and Jan~J. Sbierski.
\newblock Predictability of subluminal and superluminal wave equations.
\newblock {\em Commun.\ Math.\ Phys.}, 368:585--626, 2019.

\bibitem{ChMGHD}
Piotr~T. Chru\'{s}ciel.
\newblock On maximal globally hyperbolic developments.
\newblock {\em Jour.\ Fixed Point Theory Appl.}, 14:325--35, 2013.
\newblock arxiv:1112.5779 [gr-qc].

\bibitem{RingstroemBook}
Hans Ringstr{\"o}m.
\newblock {\em The {C}auchy problem in general relativity}.
\newblock ESI Lectures in Mathematics and Physics. European Mathematical
  Society (EMS), Z\"urich, 2009.

\bibitem{SCC}
Piotr~T. Chru\'{s}ciel.
\newblock {\em On Uniqueness in the Large of Solutions of {E}instein Equations
  (``{S}trong {C}osmic {C}ensorship'')}.
\newblock Australian National University Press, Canberra, 1991.




\end{thebibliography}
\end{document}